\documentclass[aip,jap,floatfix,preprint,
 amsmath,amssymb,
]{revtex4-1}
\usepackage{graphicx}
\usepackage{dcolumn}
\usepackage{bm}
\begin{document}
\title{Polarization-mediated Debye-screening of surface potential fluctuations in dual-channel AlN/GaN high electron mobility transistors}
\author{David A. Deen} \email{david.deen@alumni.nd.edu}\author{Ross Miller}\author{Andrei Osinsky}
\affiliation{Agnitron Technology, Eden Prairie, MN 55346}
\author{Brian P. Downey}\author{David F. Storm}\author{David J. Meyer}\author{D. Scott Katzer}\author{Neeraj Nepal}
\affiliation{Electronics Science and Technology Division, Naval Research Laboratory, Washington D.C. 20375}
\begin{abstract}
A dual-channel AlN/GaN/AlN/GaN high electron mobility transistor (HEMT) architecture is proposed, simulated, and demonstrated that suppresses gate lag due to surface-originated trapped charge. Dual two-dimensional electron gas (2DEG) channels are utilized such that the top 2DEG serves as an equipotential that screens potential fluctuations resulting from surface trapped charge. The bottom channel serves as the transistor's modulated channel. Two device modeling approaches have been performed as a means to guide the device design and to elucidate the relationship between the design and performance metrics. The modeling efforts include a self-consistent Poisson-Schrodinger solution for electrostatic simulation as well as hydrodynamic three-dimensional device modeling for three-dimensional electrostatics, steady-state, and transient simulations. Experimental results validated the HEMT design whereby homo-epitaxial growth on free-standing GaN substrates and fabrication of same-wafer dual-channel and recessed-gate AlN/GaN HEMTs have been demonstrated. Notable pulsed-gate performance has been achieved by the fabricated HEMTs through a gate lag ratio of 0.86 with minimal drain current collapse while maintaining high levels of dc and rf performance.
\end{abstract}
\maketitle
\section{Introduction}
Performance degradation due to surface and bulk charge trapping have proven to be difficult issues to resolve for high electron mobility transistors (HEMTs) derived from the III-Nitride material family.\cite{Binari,Vetury} The predominant issues include current collapse\cite{Wang,Meneghesso}, knee voltage walkout\cite{Arehart}, dc-RF frequency dispersion due to virtual gate extension\cite{Roff,Coffie}, gate and drain lag\cite{Mitrofanov,Kusmik}, and power slump.\cite{Tan,Hwang} These deleterious effects stem, in part, from a high density of uncompensated surface states that impose channel depletion with slow temporal response to applied gate voltage. Bulk trapping is also present and plays an additive role in these processes. The formation of trap states that occur at the terminal growth surface are due to free atomic bonds from the abrupt termination of the periodic lattice. However, these surface states are also a critical component that, coupled with polarization fields, mediates mobile charge migration in the formation of the two-dimensional electron gas (2DEG) channel, and thus, a necessary element to the complete picture of polarization-induced 2DEGs.\cite{Ibbetson} A myriad of techniques have been explored to mitigate dc-RF frequency dispersion and its accompanying issues. Post-growth dielectric-based passivation has been the most aggressively developed method to address the deleterious effects of surface trap states and include conformal oxide and nitride depositions.\cite{Medjdoub,Lee,Huang,Koehler,Pei,Lee} This is due to the ease of processing as well as its historical success with mature technologies such as GaAs HEMTs\cite{Chou}. In a separate approach, surface chemical treatments have been investigated to minimize the effects of virtual gating on frequency performance.\cite{Meyer,Wang2,Liu} Epitaxial growth methods have been investigated with some success such as extended GaN cap layers grown in the access region as a means to decouple the charging surface\cite{Shen,Coffie,DasGupta} and p-doped cap layers to take advantage of electrically screening the field from the charged surface.\cite{Palacios} Several studies have reported on the formation of surface states in AlGaN barriers during the high temperature anneal process responsible for alloying the metallic ohmic contacts\cite{Gonzalez,Higashiwaki}. Based upon this premise, reports of regrown ohmic contacts have shown that by avoiding the high temperature anneal process, the surface state density can be reduced.\cite{Xing1,Grace1}

Despite the successes post-growth passivation techniques have had with reducing the deleterious effects of surface trapped charge, the present technological thrust for extremely high-frequency HEMT operation is in a direction that seeks to reduce lateral transistor dimensions in order to minimize electron transport times. This invariably requires vertical dimensions to simultaneously be reduced in order to maintain sufficient charge control of the 2DEG channel. This mode of advancement is the principle underlying motivator behind Moore's law for Si and compound semiconductor technologies alike. As vertical dimensions are down-scaled and the transistor's conduction channel is engineered to be closer to the terminal growth surface, carrier transport in the channel exhibits an increased sensitivity to its neighboring surface. In order to mitigate against the exacerbated effects of surface trapping on 2DEG transport in ultra-shallow channel GaN HEMTs, novel design approaches will need to be sought. 

\begin{figure}
\includegraphics[width=0.95\columnwidth]{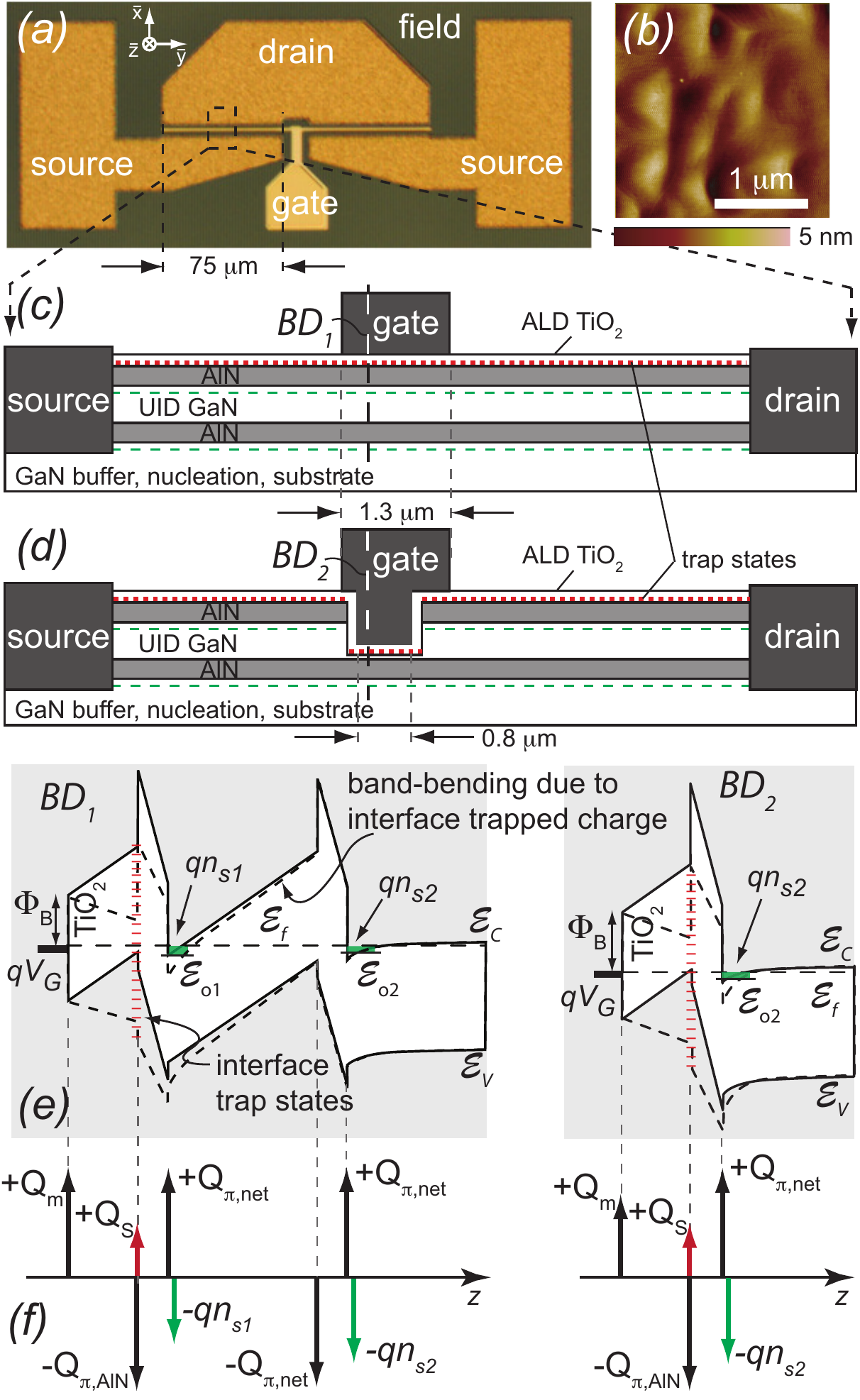}
\caption{(a) Plan-view micrograph of the HEMT topography with corresponding cross-sectional illustrations of the (c) dual-channel HEMT and the (d) recessed-gate HEMT (e) simulated band diagrams under equilibrium conditions with (dashed lines) and without (solid lines) 6$\times$10$^{13}$ cm$^{-2}$ interfacial trapped charge at the oxide/AlN interface. Interfacial trap states are depicted by red dashes in the device cross-sections and band diagrams. Sub-band energies for the upper and lower quantum wells are depicted in the illustration as $\mathcal{E}_{o1}$ and $\mathcal{E}_{o2}$, respectively, and the 2DEG charge densities of the same respective wells are given by $n_{s1}$ and $n_{s2}$. The corresponding charge distributions are illustrated in (f) where $Q_{\pi,net}$ is the the fixed net polarization charge at the AlN/GaN interface, $Q_{\pi,AlN}$ is the polarization charge of the strained AlN layer, $Q_{ns}$ is the 2DEG sheet charge density for the corresponding channels, and $Q_{T}$ is the interfacial trapped charge. A self-consistent 1D Poisson-Schrodinger solution was used to generate the band diagrams.\cite{Snider} BD$_1$ corresponds to the gated intrinsic region of the HEMT with two coincident channels and BD$_2$ corresponds to the gated intrinsic region of the recessed-gate HEMT where only a single channel remains. (b) Surface morphology of the as-grown AlN surface with a RMS roughness of 0.64 nm is shown by a $2\times2$ $\mu m^2$ AFM scan.}
\label{fig:Xsection}
\end{figure}

The AlN/GaN heterostructure affords the largest polarization fields and polarization-induced 2DEG density (up to 6$\times$10$^{13}$ cm$^{-2}$) with high mobility (1800 cm$^2$/Vs) of any III-Nitride heterostructure.\cite{Medjdoub,Zimmermann,Shinohara1, Deen1} The thickness range of the pseudomorphic AlN barrier is strain-limited to less than 5 nm \cite{Jena1,Deen1} and with the wide energy band gap (6.2 eV) and conduction band offset of the AlN layer, HEMTs fabricated from this binary-barrier heterostructure have recently set remarkable performance benchmarks for extremely high-current handling capability at high frequency\cite{Medjdoub,Zimmermann,Shinohara1,Corrion} and millimeter-wave power performance.\cite{Deen2} Yet only a handful of HEMT designs have leveraged a few of the attributes that are inherent to this particular heterostructure.\cite{Shinohara1, Shinohara2,Cao3, Deen3, Deen4,Jena2} In this report we propose and experimentally demonstrate an epitaxial-grown alternative to post-growth surface passivation based on a dual-channel AlN/GaN/AlN/GaN heterostructure through a multi-faceted assessment of its operational performance. The upper AlN/GaN heterojunction undergoes a recess etch, conformal oxidation, and gate metal deposition as illustrated in Fig. \ref{fig:Xsection}. The upper polarization-doped 2DEG serves solely to screen the potential fluctuations generated by surface trapped charge that would otherwise impose channel depletion leading to current collapse and gate lag. The trapped charge can also act as a source of remote ionized impurities that can scatter mobile channel electrons leading to mobility reduction in the current-carrying channel\cite{Deen1}, The bottom 2DEG serves as the gate-modulated channel. The HEMT access region includes both channels. Therefore, purely dual-channel AlN/GaN/AlN/GaN HEMTs have also been fabricated on the same wafer, serving as both a calibration structure for $CV$ and $IV$ characterization as well as a proxy to the recessed-gate HEMT access region. Several reports have been made on nitride-based dual-channel HEMTs with AlGaN or AlInN barriers with the intent to increase drain current density or assess HEMT noise characteristics and subsequently disregarded pulsed-gate and large-signal performance.\cite{Chu,Jha,Zhang} A notable attribute of using the AlN/GaN heterostructure for the HEMT design reported in this work is that the AlN barrier layers are inherently thin ($<$ 5 nm), which allows extremely shallow channels and therefore, multiple channel designs to maintain channels in close proximity to the surface as a means to boost drain current density without sacrificing gate charge control. This is not the case for ternary barriers that require a sufficient thickness in order to induce 2DEG formation.

The organization of this manuscript follows first with a discussion of the pertinent elements of the quantitative simulations in section \ref{section:SF}. Electrostatic simulations were made through a one-dimensional self-consistent Poisson-Schrodinger solution\cite{Snider} for calculating band diagrams, subband energies, band offsets, and other pertinent qualities of the heterostructures taken vertically through the gated portion of the HEMT. Hydrodynamic HEMT simulations were performed on the proposed structures utilizing Sentarus Device Simulator for the computation of steady-state and pulsed-gate responses of the HEMT structures under various bias conditions. In all simulations multiple concentrations of trapped charge were considered. Electrical results of the experimental HEMT devices are shown in section \ref{section:Ex} and are accompanied by discussions. Experimental results were achieved through the epitaxial growth, device fabrication, and characterization of dual-channel and recessed-gate HEMT architectures. The active heterostructure followed an AlN/GaN/AlN/GaN sequence with thicknesses of 3/15/3/(\dots) nm. Simulation results are discussed within the context of experimental results and correlation is made that shows the functionality of the screening nature of the upper 2DEG. Finally, we summarize our findings in section \ref{section:sum}.

\section{Simulation Framework}\label{section:SF}
The simulation schedule utilized for computational assessment of the dual-channel HEMTs was dually faceted with the inclusion of a Poisson-Schrodinger calculation for developing the electrostatic picture and a three-dimensional hydrodynamic device simulation (Sentaurus Device Simulator) for the computation of the device characteristics on spatial and temporal scales. The one-dimensional Poisson-Schrodinger results discussed in section \ref{sec:electro} allow quantification of band offsets, confined energy levels, equilibrium charge distributions, band bending, and structure capacitance for a range of voltage under the trapped charge conditions of interest. The commercially available device simulator utilizes self-consistent solutions for Boltzmann transport, space charge effects (Poisson), and quantum confinement (Schrodinger) to calculate electrical characteristics of the three-terminal HEMTs under controlled conditions (i.e. defined terminal voltages, temperature, doping, pulse duration, etc.) on a defined non-uniform mesh. Both electrostatic and dynamic simulations were performed for multiple cases of trapped charge at the oxide/AlN interface. The specified trapped charge densities included 0, $10^{12}$, $10^{13}$, and 6$\times 10^{13}$ cm$^{-2}$. The absence of trapped charge (0 cm$^{-2}$) represents the ideal device in which potential fluctuations were extinguished from surface/interface trap states and served to benchmark the degrading effect of trapped charge. Ramanan et al. reported on similar work in which the effect of SiN passivation on pulsed-gate performance of AlGaN/GaN HEMTs was simulated through the same methodology.\cite{Ramanan} The resultant 2D potential profile, in combination with the Wentzel-Kramers-Brillouin (WKB) tunneling formulation, has provided insight through the calculation of tunneling probability and current for the dynamic population of surface traps. All simulations were performed on HEMTs based on three heterostructures. Namely, the purely dual-channel AlN/GaN/AlN/GaN HEMT, a recessed-gate dual-channel AlN/GaN/AlN/GaN HEMT, and a single-channel AlN/GaN HEMT. 

The electrostatic picture enables the conceptualization of the effect of trapped charge on the heterostructure. A self-consistent Poisson-Schrodinger solution\cite{Snider} was used to generate the one-dimensional band diagrams taken vertically through the specified regions (Fig. \ref{fig:Xsection}(e) and Fig. \ref{fig:BD}) of the heterostructure. Additional attributes included confined wave functions, subband energies, band bending, and charge profiles. The evolution of all qualities was analyzed with respect to the trapped charge condition. The conditions for comparison where the inclusion of 0 - 6$\times$10$^{13}$ cm$^{-2}$ trapped charge in increments of 1$\times$10$^{13}$ cm$^{-2}$ at the oxide/AlN interface based off previous works that have extracted similar trap state densities of the oxide/AlN junction from high-frequency capacitance-voltage ($CV$) methods and have correlated the trap density to spatially-fixed interfacial polarization states of the AlN barrier\cite{Jena2,Deen4,Ibbetson,Deen6}. The trapped charge location has been identified as characteristic to the oxide/AlN interface and simultaneously serves as a mediator of the polarization fields in the heterostructure that coincide with mobile charge accumulation at the hetero-interfaces, which form the 2DEG\cite{Jena2,Ibbetson}. Band offsets taken from Ref. 38 were used for the gated heterostructure simulations. The work function ($\Phi_B = \chi_{Ni} - \chi_{ox}$, where $\chi$ is the electron affinity of the designated material layer) used for the Ni-oxide gate was 2 eV and band offsets taken from Ref. 38 were used for the gated heterostructure simulations. Tabulated characteristics for upper and lower 2DEG density and correspondent (first) subband energy for select trapped charge densities across the plausibility range are tabulated in the inset of Fig. \ref{fig:BD}. Second subband population is feasible for the AlN/GaN quantum well with thick AlN barriers that promote larger polarization fields and higher 2DEG densities.\cite{Jena4}

\subsection{Heterostructure electrostatics}\label{sec:electro}
\begin{figure}
\includegraphics[width=\columnwidth]{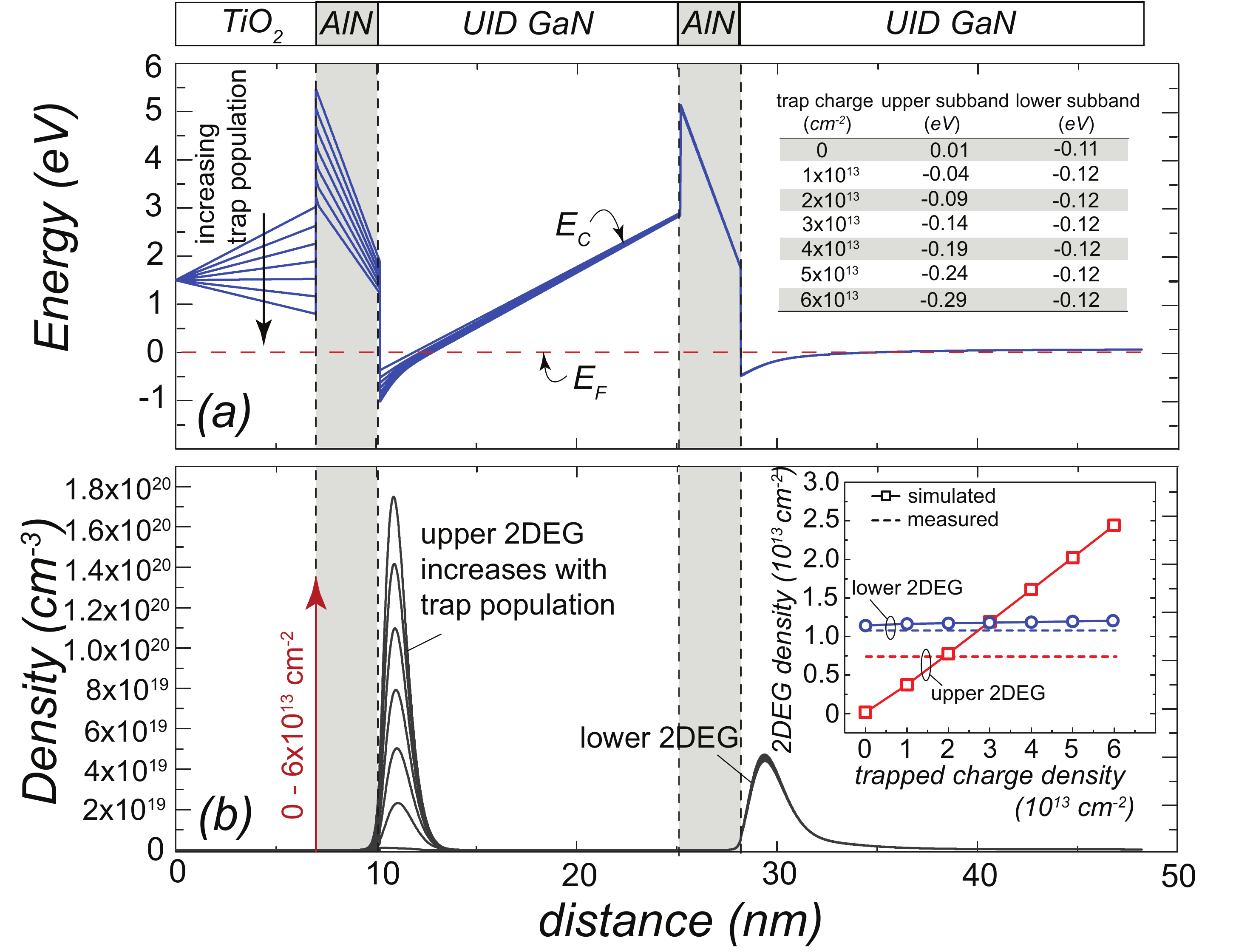}
\caption{(a) Layer structure and conduction energy band diagram and (b) corresponding charge profile of the TiO$_2$/AlN/GaN/AlN/GaN heterostructure simulated by a 1D Poisson-Shrodinger solution\cite{Snider}. Trapped charge spanning the density range of $0-6\times10^{13}$ $cm^{-2}$ was included with steps of $1\times10^{13}$ $cm^{-2}$ to evaluate its effect on both 2DEG distributions.}
\label{fig:BD}
\end{figure}

Illustrations of the dual-channel and recessed-gate HEMTs are shown in Fig. \ref{fig:Xsection} (c)-(d) along with their corresponding band diagrams in Fig. \ref{fig:Xsection} (e) to show the effect when 6$\times10^{13}$ cm$^{-2}$ trapped charge is present at the oxide/AlN interface. The correspondent sheet charge configuration is illustrated in Fig. \ref{fig:Xsection} (f). The ideal (absence of) trapped charge density as well as the single-channel heterostructure, which is representative of the gated region of the recessed-gate HEMT, is also included for contrast. In order to quantify the magnitude of the effect when charge is trapped at the oxide/AlN interface, a 2D sheet of charge was implemented to represent the trapped charge distribution and the conduction band diagram and spatial charge density are plotted in Fig. \ref{fig:BD}. Based off the simulation as well as prior work on single-channel AlN/GaN heterostructures\cite{Jena2}, the presence of (ionized positive) donor-like trap states at the oxide/AlN interface cause downward band bending, which promotes an increase in the upper 2DEG density ($qn_{s1}$) as well as the lower 2DEG density ($qn_{s2}$), though with a milder effect. This is verified in the simulated results of Fig. \ref{fig:BD}. As the donor-like traps at the oxide/AlN interface are populated (higher interfacial charge density), the conduction band is drawn downward with a commensurate increase in the upper 2DEG density. The lower 2DEG remains nearly constant during this process since the upper 2DEG is able to fully compensate the addition of interfacial charge in order to yield a net zero total charge of the heterostructure as required by charge neutrality laws. The inset in Fig. \ref{fig:BD} shows a linear increase in upper 2DEG density with trapped charge and the nearly inconsequential effect the trapped charge has on the lower 2DEG. The simulated results are plotted next to the extracted charge densities for the upper and lower 2DEGs from $CV$ profiling and shows close agreement. The response of the upper 2DEG to interfacial trapped charge supports the design principle of the recessed-gate HEMT design whereby the upper 2DEG within the access region of the HEMT is responsible for screening the potential fluctuations due to interface and/or surface trapped charge. However, it should be noted that though the 2DEG density increases under the influence of ionized surface/interface trap states, the ionized surface states trap electrons from the gate electrode and high-density 2DEG when under bias neutralizing the ionized states. This effect reverses the downward band bending and promotes channel depletion near the gate electrode where the field is the highest. The effect has been referred to as virtual gate extension. The majority of the traps have slow (dis)charge times and cannot respond to gate modulation in the range of GHz frequency or sharp gate pulses. 

\subsection{Mechanisms of interface-trap charging}
A multitude of processes exist by which surface, interface, and bulk traps populate. Some of the dominant processes include Shockley-Reed population\cite{Shockley}, Frenkel-Poole hopping, phonon-assisted trapping, and various tunneling-related population. The tunneling-type population is likely the dominant mechanism when the trap states are separated from a high-density electron gas by a barrier layer of only a few nanometers, as in the case of the AlN/GaN heterostructure. Surface and interface states may also be populated by lateral conduction from the gate electrode to states in near proximity to the drain-side of the gate. This is due to the surface potential barrier in this region being lowered by the drain bias that allows the occurrence of hopping conduction. This is the basis behind the virtual gate extension effect where electrons are trapped near the gate electrode and cannot de-trap at the same rate as the gate voltage signal, thereby serving to partially deplete the 2DEG channel within the extension region with a slow response time.
 
In this section we have given attention to trap population via quantum tunneling as the dominant mechanism for surface trap population, although other mechanisms are also at play. It is noted that thermionic processes are not accounted for in this calculation as they result in uncontrolled transmission over the barrier and amount to an upward shift in total current. A computational assessment of the dominant trapping mechanism provides insight into the efficacy of the transistor design and details of the method by which surface trapping occurs. The region of the transistor that is most susceptible to trapping/de-trapping of surface charge is that in close proximity to the drain-edge of the gate where 1) the electric field magnitude is the highest due to the applied gate-drain potential difference, and 2) an ample supply of free electrons is available for trapping, either from the 2DEG or gate electrode. In the dual-channel HEMT under consideration, the upper 2DEG serves as the free electron supply for charge trapping. The rate of change of trap occupancy is equivalent to tunneling current that populates the interface traps. For this assessment we employ the Landaur-Buttiker transmission formalism\cite{Ferry,Datta} to calculate the current that tunnels from the 2DEG to the trap center at the AlN surface (see Fig. \ref{fig:BD}). The coordinate system for all calculations is shown in Fig. \ref{fig:Xsection} such that $\overline{x}$ is laterally along the channel, $\overline{y}$ is laterally perpendicular to the channel, and $\overline{z}$ is vertically along the direction of tunneling. For a laterally isotropic crystal under the assumption of parabolic energy bands where the effective mass approximation holds, the tunneling current \cite{Ferry} is given by,
\begin{multline}\label{eq:tun1}
J_{tun}=\frac{dQ}{dt}|_x \rightarrow \\ \frac{4\pi qm^*}{(2\pi)^2\hbar^3}\int^{\infty}_{0} d\mathcal{E}_z T(\mathcal{E}_z) \int^{\infty}_{0} d\mathcal{E}_t [f_l(\mathcal{E}_z,\mathcal{E}_t) - f_r({\mathcal{E}_z,\mathcal{E}_t})]
\end{multline}
where $q$ is the elemental charge, $m^*$ is the tunneling effective mass, $\hbar$ is the reduced Plank's constant, $T(\mathcal{E}_z)$ is the transmission coefficient in the vertical direction, and $f_l(\mathcal{E})$ and $f_r(\mathcal{E})$ are the electron distribution functions on the left and right side of the tunnel barrier, respectively. The first integrand is taken over the transverse energy, $\mathcal{E}_t$, and the second integrand, $\mathcal{E}_z$ over the maximum energy range. The notation, $dQ/dt |_x$, indicates a change of charge with respect to time under transient conditions and with respect to position along the channel, $x$, since the trapped charge will have a spatial variation that correlates to the potential drop in a specific region. The 2DEG charge distribution follows a Fermi-Dirac (FD) function, $f = (1+e^\eta)^{-1}$, and $\eta = (\mathcal{E}-\mathcal{E}_f)/k_BT$ is the normalized Fermi energy within the specified region. The surface trap state distribution may take on a multitude of forms depending on the quality of the barrier growth, termination properties, and atomic species present. We consider two distribution functions here. The first being a FD distribution with the form previously stated. Inserting both FD distributions in (\ref{eq:tun1}) and integrating over transverse energy, $d \mathcal{E}_t$, the result for tunneling current may be written as a single integral over the weighting function and the transmission coefficient
\begin{multline}\label{eq:tun2}
\frac{dQ}{dt}|_x =\\ \frac{qm^*k_BT}{2\pi^2\hbar^3}\int^{\infty}_{0} d\mathcal{E}_z T(\mathcal{E}_z) ln\left(\frac{1+e^{(\mathcal{E}_F-\mathcal{E}_z)/k_BT}}{1+e^{(\mathcal{E}_F-\mathcal{E}_z-qV)/k_BT}}\right)
\end{multline}
where $V = q(\mathcal{E}^l_F - \mathcal{E}^r_F)$ is the applied voltage across the barrier due to the difference between the Fermi energies on the adjacent sides of the barrier (i.e. the terminal surface where the trap states reside and the interface where the 2DEG occurs). This result is the same as the well-known Tsu-Esaki formula.\cite{Tsu} 

The second trap distribution considered follows a discrete state distribution in energy of a single-level. Namely, the trap distribution follows a Dirac delta function, $f_l(\mathcal{E}_z,\mathcal{E}_t) = \delta((\mathcal{E}_z+\mathcal{E}_t-\mathcal{E}_f^l-\mathcal{E}_1)/k_BT)$, where $\mathcal{E}_t$ is the transverse energy component, $\mathcal{E}_f^l$ is the Fermi energy on the left side of the barrier, and $\mathcal{E}_1$ is the energy of the discrete trap state. Several reports have shown evidence of a discrete surface trap state distribution for binary and compound barriers in nitride-based heterostructures\cite{Jena2,Deen5,Deen6,Ibbetson}. The result of incorporating a single state distribution in the tunneling current follows,
\begin{multline}\label{eq:tun3}
\frac{dQ}{dt}|_x =\\ \frac{qm^*k_BT}{2\pi^2\hbar^3}\int^{\infty}_{0} d\mathcal{E}_z T\mathcal{E}_z) [\mathcal{H}(\phi)-\ln(1+e^{\eta_2})]
\end{multline}
where $\mathcal{H}(\phi)$ is the Heaviside step function, $\phi=(\mathcal{E}_F+\mathcal{E}_1-\mathcal{E}_z)/k_BT$, and $\eta_2=(\mathcal{E}_F-qV-\mathcal{E}_z)/k_BT$ are the normalized energies.

\begin{figure}
\includegraphics[width=\columnwidth]{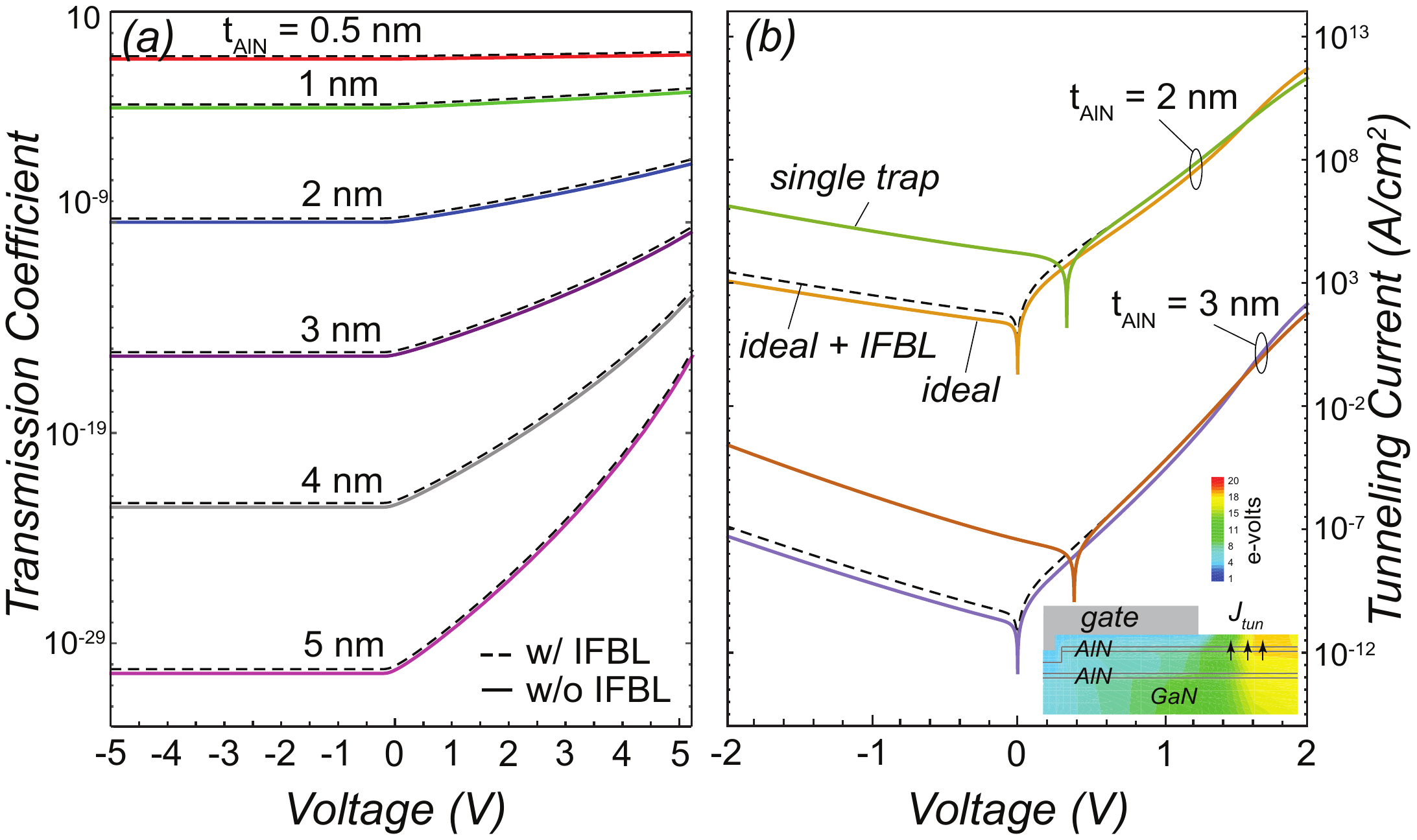}
\caption{(a) Transmission coefficient as a function of applied voltage for varied AlN barrier thickness. Dotted (solid) lines correspond to the inclusion (exclusion) of image force barrier lowering. (b) Tunneling current, $J_{tun} = dQ/dt$, as a function of applied voltage for varied AlN barrier thickness. Charge tunneling from the proximal 2DEG acts as the dominant mechanism for surface trap population under initial transient conditions. Inset shows a cross-section of the hydrodynamic calculated two-dimensional potential distribution of the recessed-gate HEMT.}
\label{fig:tunnel}
\end{figure}
The transmission function is exactly soluble for the simplistic case of a square barrier under low-field\cite{Davies,Ferry} but is not sufficiently realistic for the heterosystem being considered. The WKB approximation allows for the inclusion of realistic physics (image force barrier lowering, multiple barriers, etc.) while maintaining high computational accuracy. The WKB approximation for the transmission coefficient in the vertical direction takes the form,
\begin{equation}
T(\mathcal{E}_z) = e^{-2\int_{t_{bar}} \kappa(z) dz}
\end{equation}
where $\kappa(z) = \sqrt{2m^*(V_o(z)-\mathcal{E}(z))}/\hbar$ is the spatially-dependent wave vector inside the barrier region with the barrier potential energy $V_o(z)$, $\mathcal{E}(z)$ is the incident electron energy, and all other parameters as previously defined. The wave vector, $\kappa(z)$ is integrated over the barrier thickness, $0 < z < t_{bar}$, occurring in the argument of the exponential. Through the inclusion of image force barrier lowering (``\emph{IFBL}''), the barrier potential is reduced with a $z^{-1}$ dependency. Therefore, the AlN barrier potential energy is given by, $V(z) = q\Phi_B - qzF_{AlN} - q^2/16\pi \epsilon z$, where the last term accounts for barrier lowering. The field in the AlN barrier layer is dictated by the polarization charge and 2DEG density, $F_{AlN}=q(Q^{\pi}_{AlN}-qn_s)/\epsilon_r\epsilon_o$, where $Q^{\pi}_{AlN}$ is the polarization charge at the AlN surface and the gate voltage dependence occurs through the $n_s$ term. 

The transmission coefficient versus applied voltage for the range of AlN thicknesses where pseudomorphic strain does not impose lattice relaxation is plotted in Fig. \ref{fig:tunnel}(a). The voltage range chosen corresponds with the potential energy distribution in the proximity to the drain edge of the gate as given by the 2D hydrodynamic solution for the dual-channel structure as depicted in the inset of Fig. \ref{fig:tunnel}(b). The calculation includes the transmission response with (dotted lines) and without (solid lines) image force barrier lowering along with band gap shrinkage due to the effect of lattice spacing reduction on the band structure of the pseudomorphically strained AlN layer \cite{Ridley}. The thickness dependence on transmission coefficient is shown by a reduced coefficient as the AlN thickness increases. Positive voltage enhances electron transmission due to effective barrier thinning when the field is strong enough to promote a Fowler-Nordheim-type tunneling process.

Tunneling current is calculated by convolution of the transmission coefficient and the distribution function difference (Eqs. \ref{eq:tun2} and \ref{eq:tun3}). The voltage dependence of tunneling current is shown in Fig. \ref{fig:tunnel}(b) for 2 nm and 3 nm AlN thicknesses. The calculations included three conduction scenarios. Namely, no energy barrier lowering and both distributions of the FD type (``\emph{ideal}''), inclusion of IFBL (``\emph{ideal + IFBL}''), and a single trap state distribution with the inclusion if IFBL (``\emph{single trap}''). The latter condition is expected to be the most realistic based off prior work and the trap state is located at 0.65 eV below the conduction band edge in our calculation.\cite{Deen5,Ibbetson} The inset in Fig. \ref{fig:tunnel}(b) depicts a cross-section of the recessed-gate HEMT with an overlay of the 2D potential energy map resulting from the hydrodynamic simulations that serve to define the voltage range of interest for the tunneling current calculations. Inclusion of IFBL causes a slight increase in tunneling current as depicted by the dotted lines. A more pronounced increase in tunneling current is observed for the case when a single trap distribution occurs on the terminal (top) side of the barrier. For this condition tunneling current increases several orders of magnitude for negative bias whereas for positive bias it shows parity with the ideal + IFBL condition. Additionally, an increase in 1 nm of AlN thickness causes exponential suppression of tunneling current as shown between the comparison between 2 nm and 3 nm AlN. Therefore, we conclude that 3 nm AlN thickness optimizes the balance between high 2DEG density, mobility, and tunneling to surface trap states, and that surface trapping is enhanced when a high density state is present within the energy range accessible by the voltages imposed on the heterostructure. 

\subsection{Hydrodynamic transport simulation}
\begin{figure}
\includegraphics[width=\columnwidth]{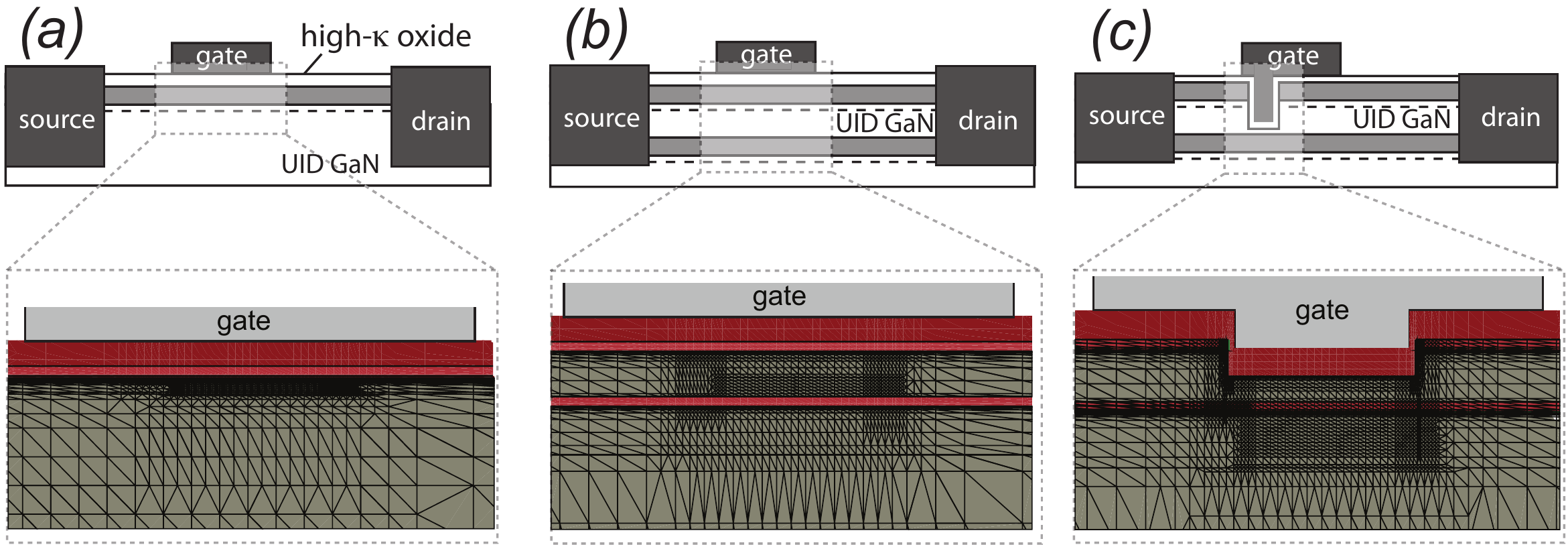}
\caption{Mesh topography in the gated region of the HEMTs for the FEA calculations used for device simulations for the (a) single-channel HEMT, (b) dual-channel HEMT, and the (c) recessed-gate HEMT.}
\label{fig:mesh}
\end{figure}

Physical simulation of transistor electrical characteristics have been performed by the commercial software Synopsis Sentaurus. Hydrodynamic (energy conserving) simulations employ a finite element mesh that can be seen in Fig. \ref{fig:mesh} and calculate quantum confinement (Schrodinger), space charge effects (Poisson), and Boltzmann transport across the three-dimensional device structure. Simulations were performed on three device structures, 1) an insulated-gate dual-channel AlN/GaN/AlN/GaN HEMT with active layer structure thickness following 3/15/3/(\dots) nm, 2) a top-channel recessed-gate HEMT with an access region heterostructure of AlN/GaN/AlN/GaN and a heterostructure below the gate that follows AlN/GaN with thicknesses of 3/(\dots) nm, 3) a single channel AlN/GaN baseline with corresponding thicknesses of 3/(\dots) nm. 

\begin{figure*}
\includegraphics[width=\textwidth]{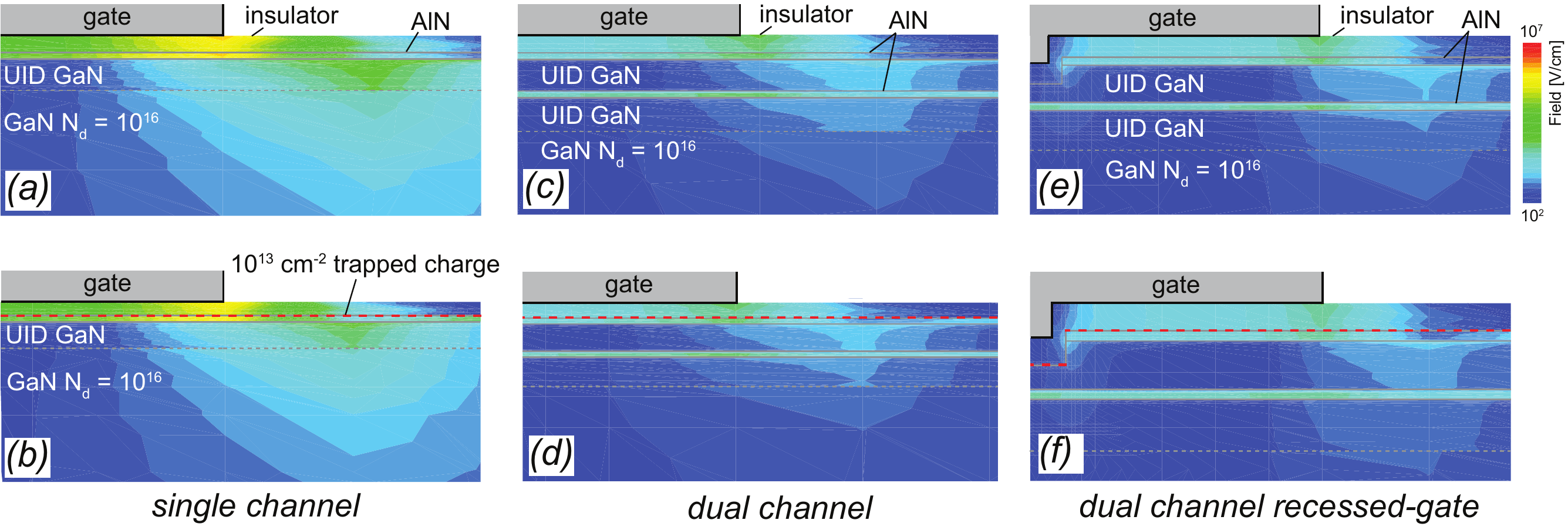}
\caption{Electric field (magnitude) distributions throughout two-dimensional cross sections in the vicinity of the drain-edge of the gate of (a-b) the single-channel HEMT, (c-d) the dual-channel HEMT, and (e-f) the recessed-gate HEMT architectures. The top row (a, c, e) shows the three HEMT designs under ideal conditions by excluding fixed interfacial charge at the oxide/AlN interface. The bottom row (b, d, f) shows the same designs when interfacial fixed charge is implemented at an areal density of $10^{13}$ cm$^{-2}$. All HEMTs are biased in saturation with forward (positive) gate bias. Color gradients indicate the variation of electric field magnitude throughout the cross-section in the region of maximum electric field.}
\label{fig:EMag}
\end{figure*}

All structures included a 7 nm thick high-k ($\epsilon_r$ = 10) gate insulator for gate leakage current reduction and a background impurity doping of 10$^{16}$ cm$^3$. Gate length was defined as 200 nm long with a width of 150 $\mu$m. Contact resistance, $R_C$, was set to 2 $\Omega$-mm based off of prior work.\cite{Deen4} The definition of the finite element mesh for the three device structures investigated is shown in Fig. \ref{fig:mesh}. In order to calibrate and establish initial operating conditions, electrostatic simulations were initially carried out. The structures were simulated under two conditions, the first being in ``ideal mode'' where interfacial trapped charge was excluded from the device. This provided a picture of the maximum operating capability of the devices as a benchmark. The second condition used in the electrostatic model was a fixed interfacial charge density of 10$^{11}$ - 10$^{13}$ cm$^{-2}$ with steps of 10$^{12}$ cm$^{-2}$ implemented at the oxide/AlN interface in the gate-drain access region. The purpose of the fixed 2D interfacial charge was to serve as a proxy for filled traps in order to observe their effects on transistor characteristics. 

\begin{figure}
\includegraphics[width=\columnwidth]{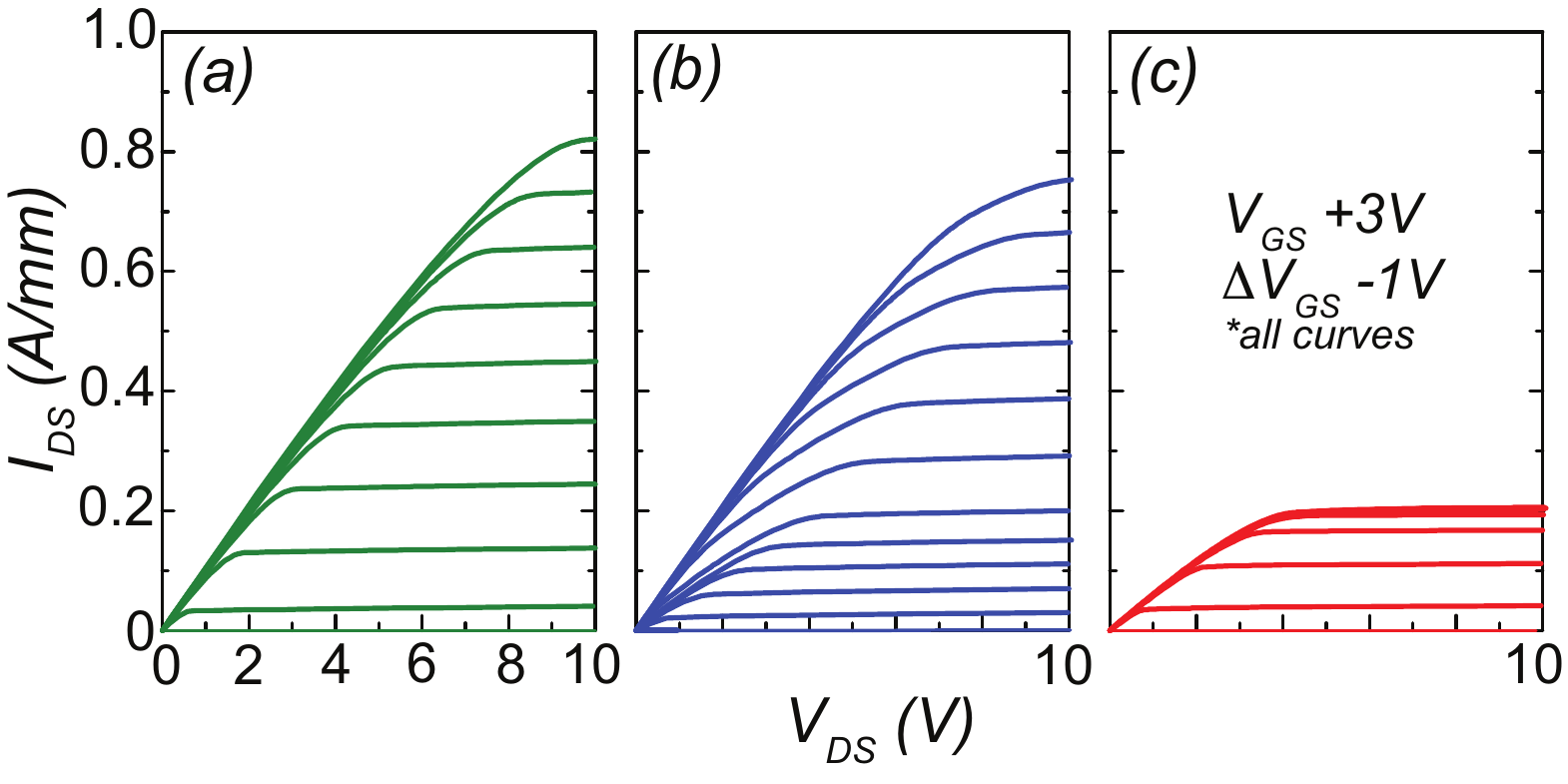}
\caption{Simulated family of drain characteristics for the (a) single-channel, (b) dual-channel, and (c) recessed-gate HEMTs. Gate-source bias conditions are listed in the figure.}
\label{fig:IVfamily}
\end{figure}

\begin{figure}
\includegraphics[width=\columnwidth]{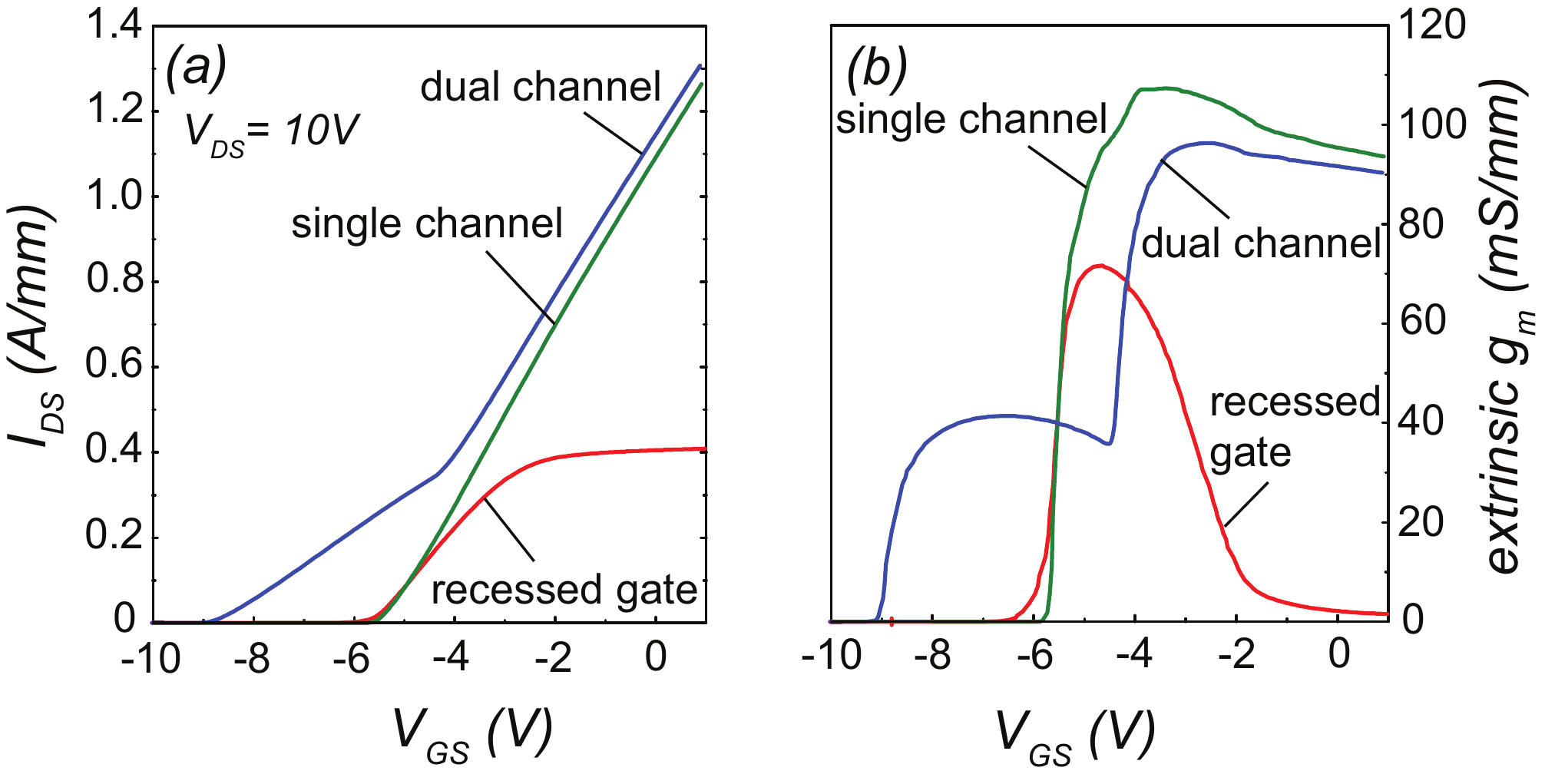}
\caption{Simulated (a) transfer and (b) transconductance characteristics for the single-channel, dual-channel, and recessed-gate HEMTs with a drain-source bias of 10V.}
\label{fig:XferSim}
\end{figure}
Two-dimensional device cross-sections with overlay maps of electric field magnitude are shown in Fig. \ref{fig:EMag} for both the ideal structures without interfacial traps and with interfacial traps implemented at a density of 10$^{13}$ cm$^{-2}$. Two points can be made from the cross-sections, first, the electric field magnitude is diminished near the drain edge of the gate for the dual-channel and recessed-gate architectures compared with the single-channel baseline and owing to the upper 2DEG channel that serves as the screening structure. Second, if one compares the distribution in electric field magnitude for the recessed-gate HEMT without and with interfacial trapped charge, it can be seen that there is only minor change between the two cases but the change that does exist shows the field magnitude to be reduced in the vicinity of the gate when trapped charge is present. This supports the premise that the upper 2DEG channel operates as an electrostatic screening layer from interfacial charge above it.

Simulated drain characteristics are shown in Fig. \ref{fig:IVfamily} and transfer and transconductance characteristics are shown in Fig. \ref{fig:XferSim}. The pertinent electrode bias voltages are listed in the figures and apply to all structures. The recessed-gate HEMT showed the  lowest maximum drain current due to enhanced access resistance resulting from channel depletion that arose from the added GaN and AlN layers that introduce additional polarization fields. Transfer characteristics showed the dual-channel HEMT to have two distinct slopes as a result of the two channels present below the gate. This resulted in a doubled-peaked transconductance that can be seen in Fig. \ref{fig:XferSim}(b). The recessed-gate HEMT showed a threshold voltage equivalent to the single-channel HEMT but with a reduced maximum current, in agreement with the drain characteristics.

\begin{figure}
\includegraphics[width=\columnwidth]{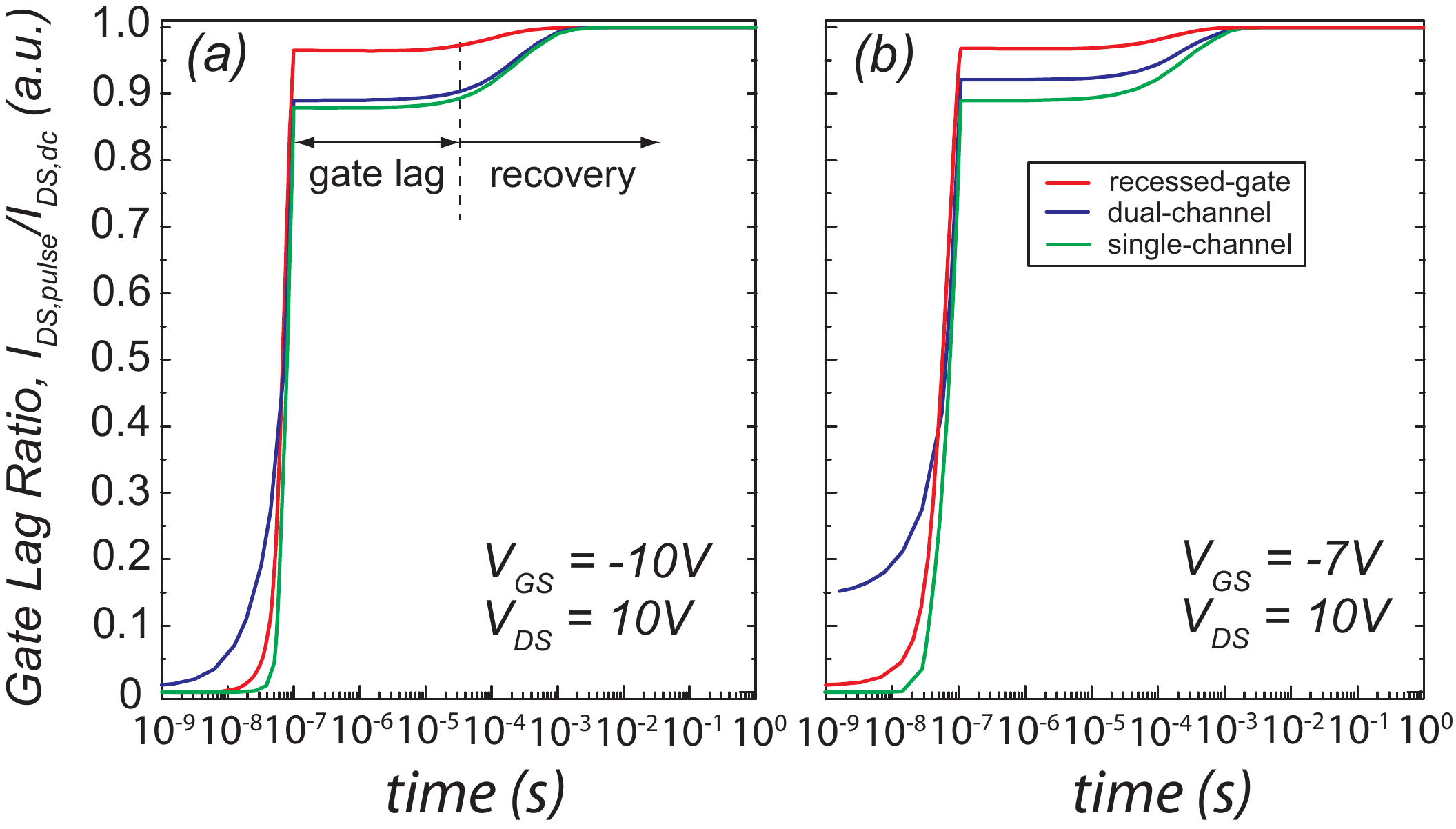}
\caption{Pulsed gate-lag response for the three HEMT architectures under investigation. (a) Gate lag results for a bias-stress of $V_{GS}$ = -10V and (b) $V_{GS}$ = -7V.}
\label{fig:GLsim}
\end{figure}

The final stage in the simulation schedule was to incorporate dynamic (voltage and time dependent) interfacial traps at the oxide/AlN interface throughout the entire device in all device structures and simulate the pulsed transient response of the three transistor architectures under these conditions. The trap dynamics were characterized through carrier tunneling and SRH-type population mechanisms. In the simulation schedule the device was initially voltage-stressed in the subthreshold regime for several minutes, then the gate voltage was switched on and the recovery of drain and gate current was monitored with respect to time. The trap states were imposed at the oxide/AlN interface and extended laterally throughout the entire device as depicted in Figs. \ref{fig:Xsection} and \ref{fig:EMag}. The trap state density was targeted at 4x10$^{13}$ cm$^{-2}$, which is in agreement with prior reports\cite{Jena2,Deen5} and at an energy level 0.65 eV below the conduction band edge of the AlN. It is also noted that no other trap states were included such as bulk GaN traps, AlN barrier traps, or oxide traps. Excluding other traps allows the simulation to demonstrate the effect from solely the interface, which is taken to be the dominant trap location that causes the impairments of pulsed and large signal responses for these ultra-thin barrier HEMT architectures. Within this framework the simulated gate lag ratio (GLR = $I_{DS,pulse}/I_{DS,dc}$) represents degradation due only to surface-originated trapping. However, bulk traps are expected to be present in physical device structures and serve to further degrade pulsed and large signal response to some extent depending on their characteristics (energy distribution, density, spatial distribution, etc.). Figure \ref{fig:GLsim} shows the simulation results for pulsed-gate response of the transistor architectures at two different intensities of gate-source voltage stress. Recovery time from the leading edge transient pulse occurs after $\sim$ $10^{-3}$ seconds whereby drain current returns to its dc reference value. As is evident from the plotted response, gate lag occurs for all HEMT structures under test but shows a diminished severity for the recessed-gate HEMT architecture due to the screening nature of the upper channel.  

\section{Experimental Results}\label{section:Ex}
AlN/GaN heterostructures were grown by RF-plasma assisted molecular beam epitaxy (MBE) on free-standing hydride vapor phase epitaxy (HVPE) GaN substrates. All heterostructures were grown at 650$^o$ C. The GaN layers were grown in the metal-rich regime. The AlN layers were grown nearly stoichiometrically. All layers were grown continuously without interrupts. Growth was initiated by a 60 second nitridation of the HVPE GaN substrate surface, immediately followed by growth of an ultra-thin, 1.5 nm AlN nucleation layer.\cite{Cao3} Next, a 1.3 $\mu$m thick 10$^{19}$ cm$^{-3}$ beryllium-doped GaN layer was deposited followed by a 0.3 $\mu$m thick lesser-doped region (2$\times$10$^{17}$ cm$^{-3}$)\cite{Deen1,Deen4}. It has been found that thick GaN:Be layers are effective for obtaining low buffer leakage in HEMT structures on native GaN substrates.\cite{Storm2,Storm} Next, a 200-nm unintentionally-doped GaN buffer layer was grown. The active heterostructure layers were subsequently grown following an AlN/GaN/AlN sequence with correspondent thicknesses of 3/15/3 nm. These layer thicknesses were chosen in order to avoid lattice relaxation of the strained AlN layers while maintaining the optimal $\mu$-$n_s$ product\cite{Deen1,Jena1} as well as following the structure investigated in the modeling portion of this work. Post-growth characterization by atomic force microscopy showed (Fig. \ref{fig:Xsection}) a surface roughness of 0.64 nm in a 2$\times$2 $\mu$m$^2$ scan without indication of lattice relaxation of the AlN layers. An inductive-based contact-less sheet resistance measurement showed as-grown room-temperature sheet resistance across the wafer to be 340 $\Omega/\Box$ indicating the population of one or both 2DEG channels in the as-grown structure.

An ohmic-first processing schedule was employed to ensure the best conditions for forming low-resistance ohmic contacts to both parallel 2DEG channels.\cite{Deen3,Deen4} The recessed-gate design in Fig. \ref{fig:Xsection} does not require low-resistance ohmic contacts to the upper 2DEG channel. However, due to the same-wafer processing of both device structures contacts were made to both 2DEG channels. A pre-metallization Cl-based dry etch was employed to etch through the top AlN and GaN layers prior to contact metallization. The target etch depth was 18 nm below the terminal surface at the interface made between the GaN spacer and the bottom AlN layer. Electron beam deposition was used to deposit a Ti/Al/Ni/Au metallic layer structure with corresponding thicknesses of 30/200/40/20 nm. An 860$^o$ C rapid thermal anneal was performed for 30 seconds following the metal contact deposition and resulted in a contact resistance of $\sim$4 $\Omega$mm. Mesa and inter-device isolation was facilitated by a conventional Cl-based dry etch. I-line stepper lithography was used to define gate feature with a target length of 0.7 $\mu$m on half of the HEMTs on wafer. A Cl-based dry etch was utilized for the gate recess etch. The target depth was 17 nm from the terminal surface such that 1 nm of GaN spacer remained as a way to ensure no plasma damage was incurred by the lower AlN barrier and 2DEG channel. Following the gate recess, atomic layer deposition (ALD) was used to deposit a conformal 7 nm thick film of TiO$_2$ for gate insulation. Optical lithography was also used for the definition of 1 $\mu$m gates and other large-gated test structures following oxide deposition. A Ni/Au gate metal deposition and lift-off concluded the fabrication. Pertinent transistor geometries were source-drain separation ($L_{DS}$) of 5 $\mu$m and gate width ($W_G$) of 150 $\mu$m.

Post oxidation Hall effect measurements were performed on Van der Pauw structures and included both channels. The room-temperature sheet resistance, charge density, and Hall mobility were determined to be $R_{sh}$ =  220 $\Omega/\square$, $n_s$ = 1.8 $\times$ 10$^{13}$ cm$^{-2}$, $\mu$ $\sim$ 1600 cm$^2$/Vs. The measured mobility represents an averaged mobility of the two parallel channels since there was no convenient means to differentiate between the channels with the standard on-wafer Hall measurement. However, the individual charge densities of each channel were determined through capacitance-voltage ($CV$) profiling discussed next. A drop in sheet resistance through an increase in both charge density and mobility was observed upon deposition of the ALD TiO$_2$ dielectric (as-grown Hall results were $R_{sh}$ =  340 $\Omega/\square$, $n_s$ = 1.25 $\times$ 10$^{13}$ cm$^{-2}$, $\mu$ $\sim$ 1500 cm$^2$/Vs). It is possible this drop in sheet resistance, in part from the improved mobility, is due to some alleviation of remote surface roughness scattering that is typical at low charge densities \cite{Deen1}. When the dielectric is deposited a higher 2DEG density is induced which provides more efficient screening of these scattering mechanisms and thus, a higher mobility results. This pertains only to the top channel.

\begin{figure}
\includegraphics[width=\columnwidth]{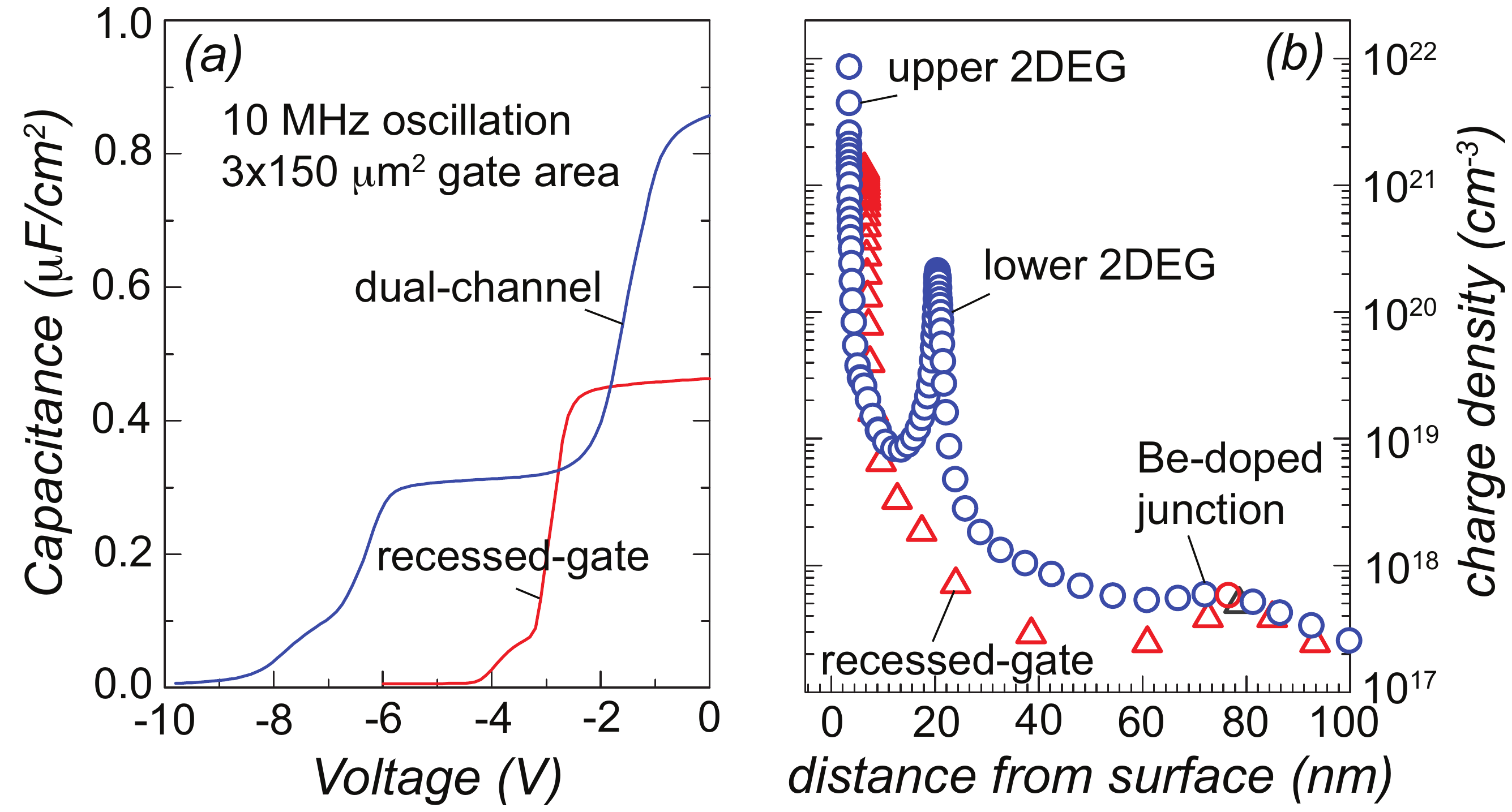}
\caption{(a) Capacitance-voltage characteristics and (b) charge profile for the dual-channel test capacitor showing two distinct capacitance plateaus corresponding to the two parallel 2DEG distributions.}
\label{fig:cv}
\end{figure}
$CV$ measurements were taken on a 100 $\mu$m diameter test capacitor with an oscillation frequency of 10 MHz and showed two distinct capacitance plateaus indicating two separate charge distributions in the heterostructure (Fig. \ref{fig:cv}a). The plot shows two curves corresponding to the pure dual-channel and the recessed-gate structure. The integration of the smaller capacitance plateau ($Q = CV$) yields a charge density of 1.08 $\times$ 10$^{13}$ cm$^{-2}$ associated with the lower 2DEG. The integration of the second capacitance plateau yields a combined charge density of 1.8 $\times$ 10$^{13}$ cm$^{-2}$ which is in agreement with the charge density measured by Hall effect. The difference of these two densities yields the upper 2DEG density and was found to be 7.25 $\times$ 10$^{12}$ cm$^{-2}$.  In both curves an inflection is shown close to bottom channel threshold. It is likely that this is a signature of the Be-doping in the GaN buffer that is employed for the buffer to be semi-insulating. The UID/Be-doped interface is $\sim$100 nm below the channel and in relatively close proximity to the 2DEG, whereby, under gate bias the Be-dopants ionize and supply a small population of positive charge in the GaN buffer. From a design standpoint this additional background charge is undesirable but can be mitigated by increasing the UID GaN layer thickness such that the Be-doped GaN is further away from the 2DEG than 100 nm. It is noted that the Be-doping does not diffuse under thermal annealing as we have observed via post-anneal secondary ion mass spectroscopy (not shown). The $CV$ data was used to calculate the approximate charge density profile by $n(z) = (C^3/q\epsilon_s)(dC/dV)^{-1}$ and is shown in Fig. \ref{fig:cv}(b). These charge densities were used to calibrate the electrostatic conditions used to calculate the band diagrams shown in Fig. 1. 

\begin{figure}
\includegraphics[width=\columnwidth]{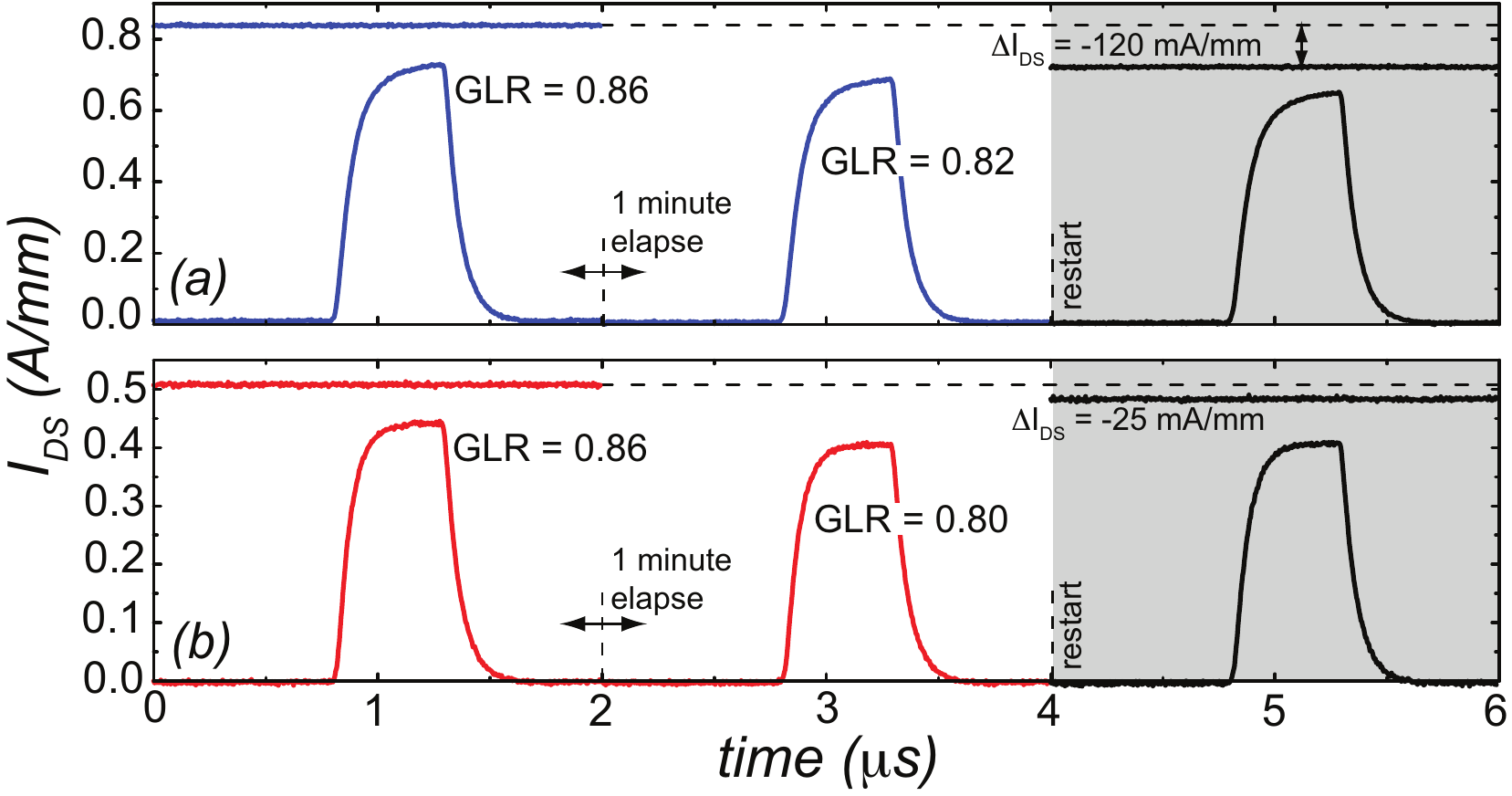}
\caption{Pulsed gate lag characteristics for the (a) dual-channel HEMT  and (b) the recessed-gate HEMT with bias conditions of $V_{DS}$ = $20V$ and $-20V < V_{GS} < 0V$ for both HEMT architectures. Pulse width for both pulse sequences was 0.5 $\mu$s.}
\label{fig:GLR}
\end{figure}
\begin{figure}
\includegraphics[width=\columnwidth]{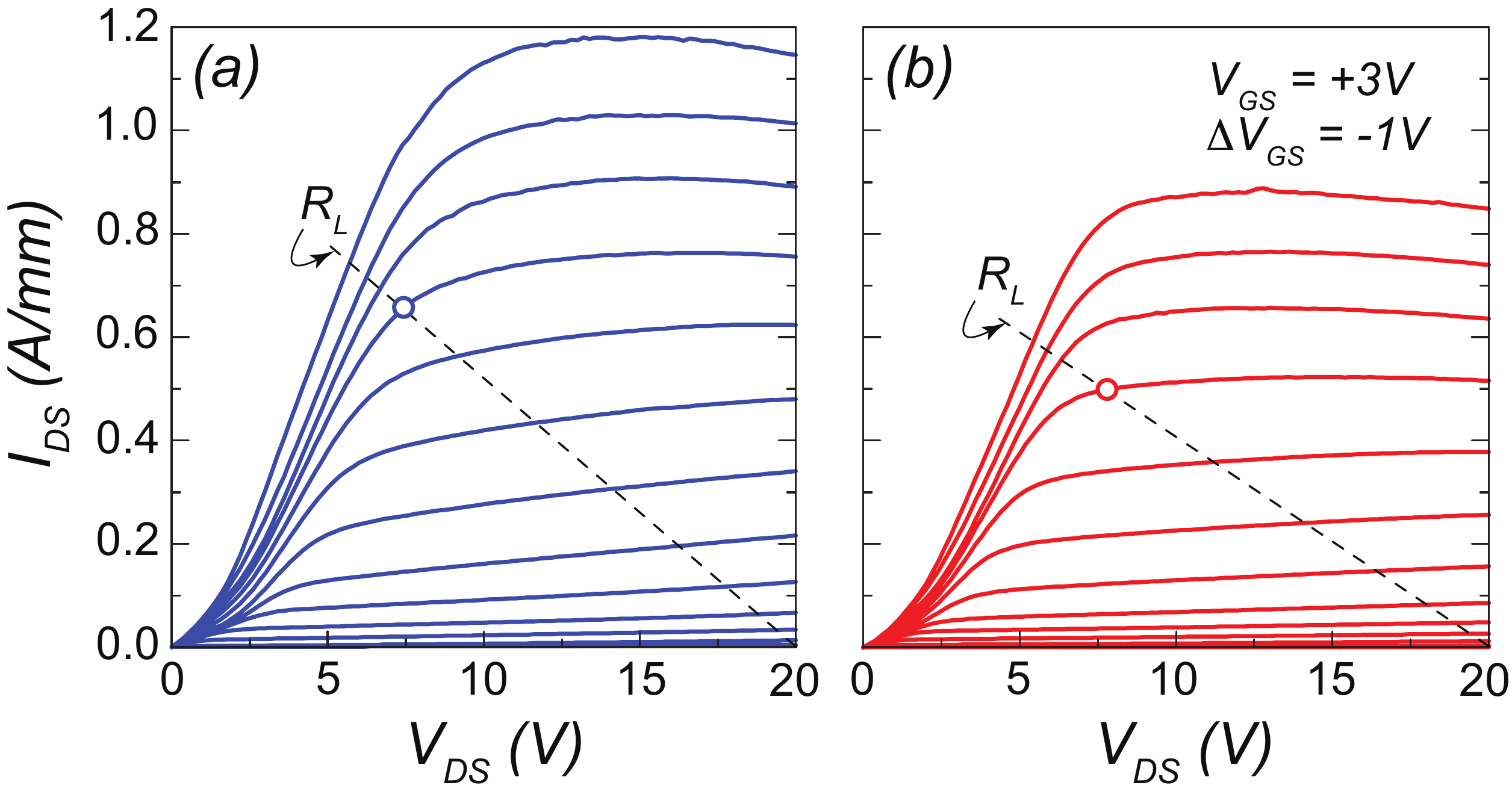}
\caption{Drain current characteristics for (a) the dual-channel HEMT and (b) the gate-recessed HEMT with corresponding load-lines (dashed) to the gate-lag measurement for a load resistance of $R_L = 277 \Omega$.}
\label{fig:drainIV}
\end{figure}

Gate lag refers to the time delay of a HEMT's drain current recovery in response to a gate voltage pulse. Gate lag results from a slow recovery from depletion of the channel charge due to proximal trapped charge.\cite{Binari,Wang} Interfacial trapped charge such as those at the oxide/AlN interface\cite{Deen4} can lead to gate lag.\cite{Binari} Therefore, a temporally-sequential pulsed gate voltage lag measurement has been used to quantify the gate lag response of the dual-channel and recessed-gate HEMTs. The measurement schedule began by measuring open-channel drain current with predetermined values of $V_{DS}$ and $V_{GS}$. Those values were $V_{GS} = 0 V$ with $V_{DS} = 10 V$ for both the dual-channel and recessed-gate HEMTs. The $V_{GS}$ values were chosen to maximize the current-voltage product across the loaded transistors while maintaining a significant gate voltage, at or above 0V (drain bias resistor value chosen to optimize the load-line based on $IV$ characteristics shown in Fig. \ref{fig:drainIV}(a) and (b)). The open-channel drain currents for the stated $V_{GS}$ values used in our measurements were $I_{DS} = 0.84$ A/mm and $0.51$ A/mm for the dual-channel and recessed-gate HEMTs, respectively (see Fig. \ref{fig:GLR}). The measured drain current density in the dc open-channel condition ($I_{DS,o}$) is then used as the normalization value when calculating the gate lag ratio as defined by $GLR = I_{DS,pulse}/I_{DS,o}$. Following the dc $I_{DS,o}$ measurement, $V_{GS}$ was brought to a value within the sub-threshold regime for a prescribed amount of time (0.8 $\mu$s), which allowed charging of trap states. Then $V_{GS}$ was abruptly pulsed back to the open-channel value previously listed for a specified pulse duration (0.5 $\mu$s) and the drain current was monitored during the pulse cycle ($I_{DS,pulse}$) before $V_{GS}$ was finally brought back into sub-threshold. In our measurement schedule shown in Fig. \ref{fig:GLR}, we additionally made successive gate pulses 1 minute apart ($V_{GS}$ held in sub-threshold between pulses) in order to observe the effects of higher trapped charge density on GLR. Moreover, our measurement schedule included a restart where all bias voltages where brought to $0 V$ immediately before repeating the measurement schedule just described. This allows for the quantification of how degraded the dc $I_{DS,o}$ value has become after pulsed bias stress (grey region in Fig. \ref{fig:GLR}) and serves as a proxy for current slump. The quantity, $\Delta I_{DS} = I_{DS,o} - I_{DS,1}$, where $I_{DS,1}$ is the dc value of $I_{DS}$ measured upon restarting the gate pulse schedule.

The results of pulsed gate lag measurements are shown in Fig. \ref{fig:GLR} for the (a) dual channel AlN/GaN HEMT and the (b) recessed-gate HEMT. The dual channel HEMT in (a) demonstrated a GLR of 0.86 which decreased while the HEMT was biased in sub-threshold to 0.82 for subsequent pulses. The hypothesis is that this reduction is mainly due to surface state charging and corresponding depletion of the upper channel (current collapse) since it was shown that the upper channel makes up a large fraction of the total drain current for the dual-channel HEMTs. Upon restarting the measurement the open channel drain current was found to have diminished by 120 mA/mm ($\Delta I_{DS}$) which is indicative of 2DEG channel depletion and possibly some buffer trapping. The recessed-gate HEMT in (b) demonstrated a GLR of 0.86-0.8, which indicates strong suppression of interface trap related gate lag degradation by its near unity value. The traps are assumed to be located at the oxide/AlN interface as is denoted in Fig. \ref{fig:Xsection}.\cite{Zhang},\cite{Jena2} The recessed-gate HEMT showed an emphasized (dis)charge curvature of the drain current pulse in Fig. \ref{fig:GLR}(b). This may be a manifestation of increased gate-to-channel capacitive charging time between the gate metal and upper 2DEG. Further design enhancements are anticipated to alleviate some of the $RC$ charging in the HEMT design. A notable result of the recessed-gate HEMT is that after the GLR sequence was stopped and restarted, the initial drain current density had not diminished ($\Delta I_{DS} = I_{DS,o} - I_{DS,1}$ = 0) despite the absence of a passivation layer other than the thin TiO$_2$ gate insulation. Although other reports have been made on nitride-based dual-channel HEMTs with alloyed barrier layers\cite{Chu, Jha, Zhang}, none have included gate lag measurements. Although not shown, we have typically observed single channel AlN/GaN HEMTs grown on sapphire or SiC substrates with comparable oxide layer thicknesses to have GLRs of $<$ 0.5. Further refinements in the contacts, gate process, and layer structure are anticipated to advance the design to fully mitigate the detriment of surface traps observed in these ultra-shallow channel AlN/GaN HEMTs.

\begin{figure}
\includegraphics[width=\columnwidth]{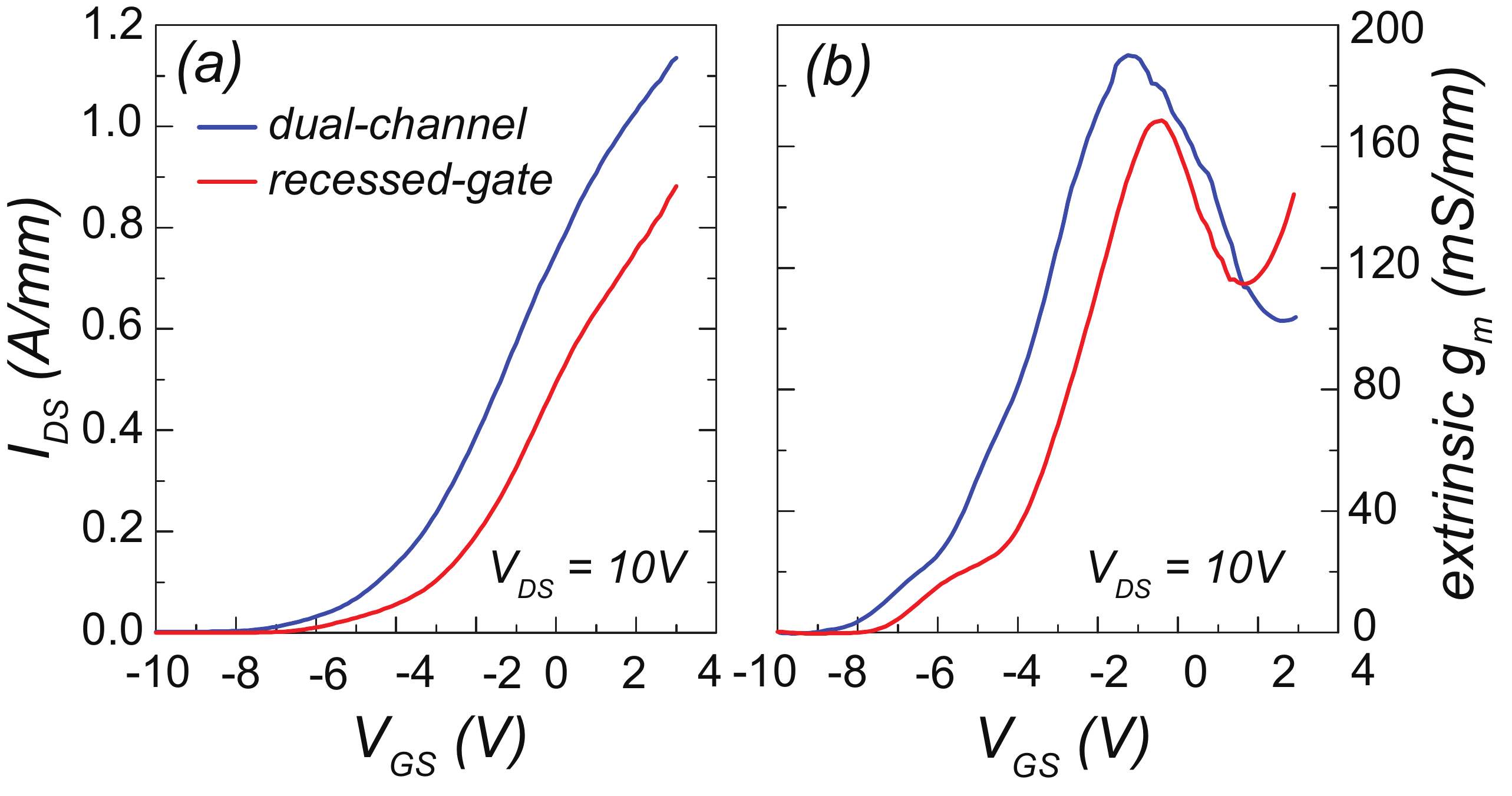}
\caption{Transfer (a) and transconductance (b) characteristics for the dual-channel and recessed-gate HEMTs with $V_{DS} = 10V$.}
\label{fig:xfer}
\end{figure}
Figure \ref{fig:drainIV} shows results for the drain characteristics of both the dual-channel HEMT (blue) and recessed-gate (red) HEMTs. The low-field drain IV curves show mild non-linearity. This indicates the metallic contacts did make pure ohmic contact to both 2DEG channels, but that the contact is a low-barrier Schottky contact. This has been observed on GaN-capped AlN/GaN heterostructures.\cite{DeenSSE} Low-resistance annealed ohmic contacts through AlN barriers have proven a challenge.\cite{DeenSSE} We anticipate improved ohmic contacts could be achieved through the employment of a recess etch and n$^+$ GaN regrowth. Nonetheless, high drain current densities of 1.2 A/mm was observed on the dual-channel HEMT and 0.9 A/mm on the recessed-gate HEMT, respectively. These maximum current values were observed at a gate voltage of $V_{GS}$ = +3V and withstood a 20V source-drain voltage. 

Transfer characteristics for both HEMT architectures are shown in Fig. \ref{fig:xfer}. The transfer characteristics showed a dual-slope indicative of the two channels as they pinched off individually. The dual-slope arises in the recessed-gate HEMT due to the gate head overlap of the top channel in the access region, which was verified by the hydrodynamic model. Consequential to the dual-sloped transfer characteristics is the presence of a peaked transconductance that follows with an inflection point. Therefore, on the gate-source voltage scale plotted, a maximum extrinsic transconductance of ~200 mS/mm was measured for the dual-channel HEMT and ~165 mS/mm for the recessed-gate HEMT. There was an observable voltage shift in the transfer and transconductance characteristics of the dual-channel HEMT as can be seen in Fig \ref{fig:xfer}. This is due to the influence of the second (top) channel to electrostatically shift the modulation of the bottom channel. 

\begin{figure}
\includegraphics[width=\columnwidth]{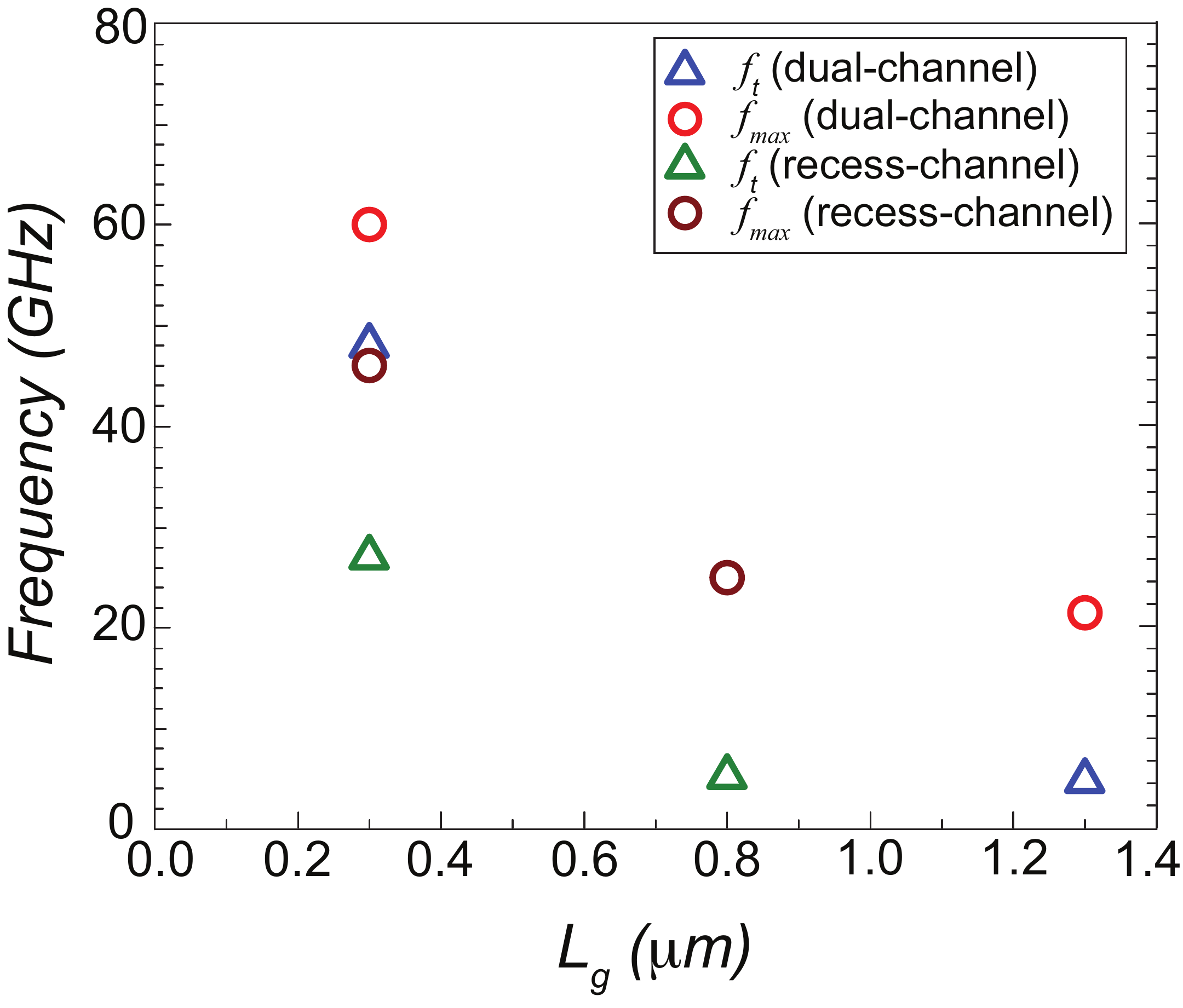}
\caption{Small signal frequency characteristics for the 1.3 $\mu$m long gates on the dual-channel HEMT, the 0.8 $\mu$m long gates on the recessed-gate HEMT, and 0.3 $\mu$m long gates on both dual-channel and recessed-gate HEMTs reported in our separate work \cite{Deen4}.}
\label{fig:freq}
\end{figure}
The recessed-gate HEMT showed peculiar transfer characteristics with an absence of a well defined maximum current density and non-linear characteristic as shown in Fig. \ref{fig:xfer}(a). A multi-peaked $g_m$ was observed as shown in Fig. \ref{fig:xfer}(b) as a result of the non-linear transfer characteristic. The non-linearity in the transfer characteristic is a manifestation of the two channels in the access region immediately beneath the T-gate head where a small portion of the dual-channel is covered in the access region (see Fig. \ref{fig:Xsection}a). Coupling between both 2DEG channels may occur, particularly for a thinner GaN spacer layers and energy barriers lower than AlN.

It is noted that in a separate but similar report we have demonstrated dual-channel and recessed-gate HEMTs with 300 nm long gates and Al$_2$O$_3$ gate insulation.\cite{Deen4} In that work similar dc and pulsed-gate $IV$ results were obtained from HEMTs with an identical heterostructure to those presented in this article. Devices reported therein achieved unity current gain frequencies, $f_t$ of 48 GHz and 27 GHz for the dual-channel and recessed-gate HEMTs, respectively. For maximum signal of oscillation, $f_{max}$, they also achieved 60 GHz and 46 GHz for the dual-channel and recessed-gate HEMTs, respectively. For completeness, those results are plotted in Fig. \ref{fig:freq} next to the small signal frequency performance metrics obtained in this work on the 0.8 $\mu$m and 1.3-$\mu$m long gated HEMTs. The resultant frequency performance demonstrates that the recessed-gate HEMT is an excellent fundamental design for a high-current amplifier or switch with the capability to operate at high-frequency.

\section{Summary}\label{section:sum}
In summary, we have proposed and demonstrated purely dual-channel and novel recessed-gate dual-channel AlN/GaN/AlN/GaN HEMT architectures that suppress surface trapped charge related gate lag. This was achieved by leveraging the upper polarization-induced 2DEG as an equipotential that screens surface potential fluctuations arising from trapped charge. Electrostatic simulations elucidated the response of the system (charge, fields, and band diagrams) to the presence of surface trapped charge and provided correlation between experimentally-derived charge densities to the calculation. Hydrodynamic three-dimensional simulations provided quantitative validation of the device architectures and elucidated the trapping picture for correlation to several pertinent physical mechanisms including charge tunneling to surface traps, transistor $IV$ characteristics, and pulsed-gate $IV$ response. Experimental proof of the conceptual HEMT designs was demonstrated through epitaxial growth, fabrication, and characterization of same-wafer dual-channel and recessed-gate dual-channel HEMT architectures. Drain characteristics showed up to 1.2 A/mm drain current for the dual-channel HEMT and 0.9 A/mm for the recessed-gate HEMT. Gate lag ratio of 0.86 was demonstrated with minimal decrease in subsequent pulses over time indicating the prohibition of current collapse. Moreover, small signal frequency performance showed HEMT capability to operate in the millimeter wave spectrum. 
\section{Acknowledgements}
The authors gratefully acknowledge fruitful technical discussion with professors Huili (Grace) Xing and Debdeep Jena at Cornell University and Hiu Wong and Nelson Garces at Synopsis for assistance with the Sentaurus simulation. The authors acknowledge N. Green at NRL for his assistance with device processing. The work at Agnitron Technology was supported by NASA and the work at NRL was supported by the Office of Naval Research (P. Maki).


\begin{thebibliography}{[1]}
\bibitem{Binari} S. C. Binari, K. Ikossi, J. A. Roussos, W. Kruppa, D. Park, H. B. Dietrich, D. D. Koleske, A. E. Wickenden, R. L. Henry, IEEE Trans. Elec. Dev., Vol. 48, No. 3, (2001).
\bibitem{Vetury} R. Vetury, N. Q. Zhang, S. Keller, U. K. Mishra, IEEE Trans. Elec. Dev., Vol. 48, No. 3, 560, (2001). 
\bibitem{Wang} M. Wang, D. Yan, C. Zhang, B. Xie, C. P. Wen, J. Wang, Y. Hao, W. Wu, B. Shen, IEEE Elec. Dev. Lett., Vol. 35, No. 11, 1094 (2014).
\bibitem{Meneghesso} G. Meneghesso, G. Verzellesi, R. Pierobon, F. Rampazzo, A. Chini, U. K. Mishra, C. Canali, E. Zanoni, IEEE Trans. Elec. Dev. 51, 1554, (2004).
\bibitem{Arehart} A. R. Arehart, A. Sasikumar, S. Rajan, G. D. Via, B. Poling, B. Winningham, E. R. Heller, D. Brown, Y. Pei, F. Recht, U. K. Mishra, S. A. Ringel, Sol. Stat. Elec. 80, 19, (2013).
\bibitem{Roff} C. Roff, J. Benedikt, P. J. Tasker, D. J. Wallis, K. P. Hilton, J. O. Maclean, D. G. Hayes, M. J. Uren, T. Martin, Trans. Elec. Dev. 56, 13, (2009).
\bibitem{Coffie} Coffie, D. Buttari, S. Heikman, S. Keller, A. Chini, L. Shen, U. K. Mishra, Elec. Dev. Lett. 23, 588, (2002).
\bibitem{Mitrofanov} O. Mitrofanov, M. Manfra,N. Weimann, Appl. Phys. Lett. 82, 4361, (2003).
\bibitem{Kusmik} Kusmik et al., Phys. Stat. Sol. A 204, 2019, (2007).
\bibitem{Tan} Tan et al., Elec. Dev. Lett. 27, 1, (2006).
\bibitem{Hwang} J. C. M. Hwang, Sol.-Stat. Elec. 43, 1325, (1999).
\bibitem{Ibbetson} J. P. Ibbetson, P. T. Fini, K. D. Ness, S. P. DenBaars, J. S. Speck, U. K. Mishra, Appl. Phys. Lett. 77, 250, (2000).
\bibitem{Pei} Y. Pei, S. Rajan, M. Higashiwaki, Z. Chen, S.P. DenBaars, U.K. Mishra, IEEE Elec. Dev. Lett. No. 4, Vol. 30, (2009).
\bibitem{Medjdoub} F. Medjdoub, M. Zegaoui, D. Ducatteau, N. Rolland, P. A. Rolland, IEEE Elec. Dev. Lett. 32, 874, (2011).
\bibitem{Lee} D. S. Lee, O. Laboutin, Y. Cao, W. Johnson, E. Beam, A. Ketterson, M. Schuette, P. Saunier, T. Palacios, IEEE Elec. Dev. Lett., Vol. 33, No. 7, 976 (2012).
\bibitem{Huang} S. Huang, Q. Jiang, S. Yang, C. Zhou, K. J. Chen, IEEE Elec. Dev. Lett., Vol. 33, No. 4, 516 (2012).
\bibitem{Koehler} A. D. Koehler, N. Nepal, T. J. Anderson, M. J. Tadjer, K. D. Hobart, C. R. Eddy, F. J. Kub, IEEE Elec. Dev. Lett., Vol. 34, No. 9, 1115 (2013).
\bibitem{Chou} Y.C. Chou, R. Lai, G.P. Li, J. Hua, P. Nam, R. Grundbacher, H.K. Kim, Y. Ra, M. Biedenbender, E. Ahlers, M. Barsky, A. Oki, D. Streit, IEEE Elec. Dev. Lett. No. 1, Vol. 24, (2003).
\bibitem{Meyer} D. J. Meyer, J. R. Flemish, J. M. Redwing, Appl. Phys. Lett. 92, 193505, (2008).
\bibitem{Wang2} R. Wang, G. Li, O. Laboutin, Y. Cao, W. Johnson, G. Snider, P. Fay, D. Jena, H. Xing, IEEE Elec. Dev. Lett. 32, 7, (2011).
\bibitem{Liu} H.-Y. Liu, B.-Y. Chou, W.-C. Hsu, C.-S. Lee, C.-S. Ho, IEEE Trans. Elec. Dev., Vol. 58, No. 12, 4430, (2011).
\bibitem{Shen} L. Shen, R. Coffie, D. Buttari, S. Heikman, A. Chakraborty, A. Chini, S. Keller, S. P. DenBaars, U. K. Mishra, IEEE Elec. Dev. Lett., Vol. 25, No. 1, 7, (2004).
\bibitem{DasGupta} S. DasGupta, L. Biedermann, M. Sun, R. J. Kaplar, M. J. Marinella, K. avadil, S. Atcitty, T. Palacios, IEEE 3C.4.1 (2013). 
\bibitem{Palacios} T. Palacios, Ph.D. Dissertation, University of California, Santa Barbara, 2006.
\bibitem{Gonzalez} F. Gonzalez-Posada, J. A. Bardwell, S. Moisa, S. Haffouz, H. Tang, A. F. Brana, E. Munoz, Appl. Surf. Sci. 253, 6185, (2007).
\bibitem{Higashiwaki} M. Higashiwaki, S. Chowdhury, B. Swenson, U. Mishra, Appl. Phys. Lett., Vol. 97, No. 22, 222105, (2010). 
\bibitem{Grace1} B. Song, M. Zhu, Z. Hu, M. Qi, K. Nomoto, X. Yan, Y. Cao, D. Jena, H. G. Xing, IEEE Elec. Dev. Let. 37, 1, (2016).
\bibitem{Xing1} Y. Yue, Z. Hu, J. Guo, B. Sensale-Rodriguez, G. Li, R. Wang, F. Faria, T. Fang, B. Song, X. Gao, S. Guo. T. Kosel, G. Snider, P. Fay, D. Jena, H. Xing, IEEE Elec. Dev. Lett., Vol. 33, No. 7, (2012).
\bibitem{Zimmermann} T. Zimmermann, D. A. Deen, Y. Cao, J. Simon, P. Fay, D. Jena, H. G. Xing, IEEE Elec. Dev. Lett. 29, 661, (2008).
\bibitem{Shinohara1} K. Shinohara, D. Regan, I. Milosavljevic, A. L. Corrion, D. F. Brown, P. J. Willadsen, C.Butler, A.Schmitz,S. Kim,V.Lee, A.Ohoka, P.M.Asbeck, and M. Micovic, IEEE Electron Device Lett., Vol. 32, No. 8, pp. 1074Ð1076, (2011).
\bibitem{Shinohara2} K. Shinohara, A. Corrion, D. Regan, I. Milosavljevic, D. Brown, S. Burnham, P.J. Willadsen, C. Butler, A. Schmitz, D. Wheeler, A. Fung, and M. Micovic, Proc. IEDM Tech. Dig., 672Ð675, (2010).
\bibitem{Deen1} D. A. Deen, D. F. Storm, D. J. Meyer, R. Bass, S. C. Binari, T. Gougousi, K. R. Evans, Appl. Phys. Lett. 105, 093503, (2014).
\bibitem{Jena1} Yu Cao and Debdeep Jena, Appl. Phys. Lett. 90, 182112, (2007).
\bibitem{Snider} I. H. Tan, G. L. Snider, L. D. Chang, E. L. Hu, J. Appl. Phys. 68, 4071, (1990).
\bibitem{Corrion} A. L. Corrion, K. Shinohara, D. Regan, I. Milosavljevic, P. Hashimoto, P. J. Willadsen, A. Schmitz, D. C. Wheeler, C. M. Butler, D. Brown, S. D. Burnham, M. Micovic, IEEE Elec. Dev. Lett., Vol. 31, No. 10, (2010).
\bibitem{Deen2} D. J. Meyer, D. A. Deen, D. F. Storm, M. G. Ancona, D. S. Katzer, R. Bass, J. A. Roussos, B. P. Downey, S. C. Binari, T. Gougousi et al., IEEE Elec. Dev. Lett. 34, 199 (2013).
\bibitem{Cao3} Y. Cao, T. Zimmermann, H. Xing, D. Jena, Appl. Phys. Lett. 96, 042102 (2010).
\bibitem{Jena2} S. Ganguly, J. Verma, G. Li, T. Zimmermann, H. Xing, D. Jena, Appl. Phys. Lett. 99, 193504 (2011).
\bibitem{Deen3} D. A. Deen, A. Osinsky, R. Miller, Appl. Phys. Lett. 104, 093506 (2014).
\bibitem{Deen4} D. A. Deen, D. F. Storm, D. S. Katzer, R. Bass, D. J. Meyer, Appl. Phys. Lett 7, 109 (2016).
\bibitem{Chu} R. Chu, Y. Zhou, J. Liu, D. Wang, K. J. Chen, K. M. Lau, IEEE Trans. Elec. Dev. 52, 4 (2005). 
\bibitem{Jha} S. K. Jha, C. Surya, K. J. Chen, K. M. Lau, E. Jelencovic, Sol.-Sta. Elec. 52, 606 (2008). 
\bibitem{Zhang} K. Zhang, J. Xue, M. Cao, L. Yang, Y. Chen. J. Zhang, X. Ma, Y. Hao, J. Appl. Phys. 113, 174503 (2013).
\bibitem{Ramanan} N. Ramanan, B. Lee, and V. Misra, IEEE Trans. Elec. Dev., Vol. 61, No. 6 (2014).
\bibitem{Deen6} D. A. Deen, D. F. Storm, R. Bass, D. J. Meyer, D. S. Katzer, S. C. Binari, J. W. Lacis, T. Gougousi, Appl. Phys. Lett. 98, 023506, (2011).
\bibitem{Jena4} Y. Cao, K. Wang, A. Orlov, H. Xing, D. Jena, Appl. Phys. Lett. 92, 152112, (2008).
\bibitem{Shockley} W. Shockley and W. T. Read, Phys. Rev. 87, 835 (1952). 
\bibitem{Ferry} D. K. Ferry and S. M. Goodnick, ``Transport in Nanostructures; $2^{nd}$ Edition'', Cambridge University Press, (2009). 
\bibitem{Datta} S. Datta, ``Electronic Transport in Mesoscopic Systems'', Cambridge University Press, (1995).
\bibitem{Tsu} R. Tsu and L. Esaki, Appl. Phys. Lett. 24, 593 (1974).
\bibitem{Deen5} D. Deen, D. Storm, D. Meyer, D. S. Katzer, R. Bass, S. Binari, T. Gougousi, Phys. Stat. Sol. C, 201001071, 1 (2011).
\bibitem{Davies} J. H. Davies, Cambridge University Press, 1998. 
\bibitem{Ridley} B. K. Ridley, W. J. Schaff, and L. F. Eastman, J. Appl. Phys. 94, 3972, (2003).
\bibitem{Storm2} D. F. Storm, D. S. Katzer, D. A. Deen, R. Bass, J. A. Roussos, S. C. Binari, T. Paskova, E. A. Preble, K. R. Evans, Sol. Stat. Elec. 54, 1470, (2010).
\bibitem{Storm} D. F. Storm, D. A. Deen, D. S. Katzer, D. J. Meyer, S. C. Binari, T. Gougousi, T. Paskova, E. A. Preble, K. R. Evans, D. J. Smith, J. Crys. Growth 380, 14, (2013).
\bibitem{Adesida} L. Wang, I. Adesida, A. M. Dabiran, A. M. Wowchak, P. P. Chow, Appl. Phys. Lett. 93, 032109 (2008).
\bibitem{DeenSSE} D.A. Deen, D.F. Storm, D.S. Katzer, D.J. Meyer, S.C. Binari, Sol. Stat. Elec. 54, 613 (2010).
\end{thebibliography}
\end{document}